\newif\ifconfver
\confverfalse      
\confvertrue        

\ifconfver
     \documentclass[10pt,twocolumn,twoside]{IEEEtran}
\else
    \documentclass[11pt,draftcls,onecolumn]{IEEEtran}
\fi

\usepackage{graphicx}
\usepackage{psfrag}
\usepackage{url}
\usepackage{stfloats}
\usepackage{amsfonts,amssymb,amsmath,bm,paralist,theorem,cite,ifthen,color}
\usepackage{array}
\usepackage{calc}
\usepackage{subfigure}
\usepackage{cases}
\usepackage{verbatim}
\usepackage{algorithm,algpseudocode}
\usepackage{multirow}
\usepackage{mathrsfs}
\usepackage{epstopdf}

{
\theorembodyfont{\rmfamily}
\theoremheaderfont{\itshape}

\newtheorem{Remark}{Remark}
}

\newtheorem{Prop}{\bf Proposition}
\newtheorem{Theorem}{\bf Theorem}

\newtheorem{Observation}{\bf Observation}
\newtheorem{Claim}{\bf Claim}
\newtheorem{Fact}{\bf Fact}

\definecolor{orange}{RGB}{255,107,0}

\newlength{\twidth}
\ifconfver
\else
\fi

    \makeatletter
    \def\multilimits@{\bgroup
  \Let@
  \restore@math@cr
  \default@tag
 \baselineskip\fontdimen10 \scriptfont\tw@
 \advance\baselineskip\fontdimen12 \scriptfont\tw@
 \lineskip\thr@@\fontdimen8 \scriptfont\thr@@
 \lineskiplimit\lineskip
 \vbox\bgroup\ialign\bgroup\hfil$\m@th\scriptstyle{##}$\hfil\crcr}
    \def\Sb{_\multilimits@}
    \def\endSb{\crcr\egroup\egroup\egroup}
\makeatother

{\begin{list}{}%
    {\setlength{\rightmargin}{0pc}%
    \setlength{\leftmargin}{1pc} }} %
{\end{list}}

{\begin{list}{}%
    {\setlength{\rightmargin}{1pc}%
    \setlength{\labelsep}{0pc}
    \setlength{\leftmargin}{1pc} }} %
{\end{list}}

{\begin{list}{}%
    {\setlength{\rightmargin}{0pc}%
    \setlength{\leftmargin}{0pc} }} %
{\end{list}}

{\begin{list}{}{
    \settowidth{\labelwidth}{\mbox{\textnormal{#1}}}%
    \setlength{\leftmargin}{\labelwidth+\labelsep}}}%
{\end{list}}

\begin{document}

\bibliographystyle{IEEEtran}

\title{Full-Duplex Bidirectional Secure Communications under Perfect  and Distributionally Ambiguous Eavesdropper's CSI}
\ifconfver \else {\linespread{1.1} \rm \fi

\author{
Qiang Li,~\IEEEmembership{Member,~IEEE}, Ying Zhang, Jingran Lin,~\IEEEmembership{Member,~IEEE,} and Sissi Xiaoxiao Wu,~\IEEEmembership{Member,~IEEE}
\thanks{Part of this work has been published in IEEE GLOBECOM 2016~\cite{zhangy16}. This work was supported in part
by the National Natural Science Foundation of China under Grants (61401073, 61531009, 61671120), and in part by the Fundamental
Research Funds for the Central Universities under Grant (ZYGX2016J011).}
\thanks{Q. Li, Y. Zhang and J. Lin are with
School of Communication and Information Engineering, University of Electronic Science and Technology of China, Chengdu, P.~R.~China. E-mails: lq@uestc.edu.cn, 201422010638@std.uestc.edu.cn, jingranlin@uestc.edu.cn}
\thanks{S. X. Wu is with the
College of Information Engineering, Shenzhen University, Shenzhen, P. R. China. E-mail: xxwu.eesissi@gmail.com.}
}


\maketitle

\ifconfver \else
\begin{center} \vspace*{-2\baselineskip}
\end{center}
\fi
\begin{abstract}
Consider a  full-duplex (FD) bidirectional secure  communication system, where two communication nodes, named Alice and Bob, simultaneously transmit and receive  confidential information from each other, and an
eavesdropper, named Eve, overhears the transmissions.
Our goal is to maximize the sum secrecy rate (SSR) of the  bidirectional transmissions by  optimizing the transmit covariance matrices at Alice and Bob.
To tackle this SSR maximization (SSRM) problem, we
develop an alternating difference-of-concave (ADC) programming approach to alternately optimize the transmit covariance matrices at Alice and Bob. We show that the ADC iteration has  a semi-closed-form beamforming solution, and is guaranteed to converge to a  stationary solution of the SSRM problem. Besides the SSRM design, this paper also deals with a robust SSRM transmit design under a  \emph{ moment-based} random channel state information (CSI) model, where only some roughly estimated first and second-order statistics of Eve's CSI are available, but the exact distribution or other high-order statistics is not known. This moment-based error model is new and different from the widely used bounded-sphere error model and the Gaussian random error model. Under the consider CSI error model, the robust SSRM is formulated as an outage probability-constrained SSRM problem. By leveraging the Lagrangian duality theory and DC programming,  a tractable safe solution to the robust SSRM problem is derived. The effectiveness and the robustness of the proposed designs are demonstrated through  simulations.
\\\\
\ifconfver
\else
\noindent {\bfseries EDICS}: MSP-CODR (MIMO precoder/decoder design), MSP-APPL (Applications of MIMO communications and signal processing), SAM-BEAM (Applications of sensor and array multichannel processing)
\fi
\end{abstract}

\ifconfver \else \IEEEpeerreviewmaketitle} \fi

\ifconfver \else
\newpage
\fi



\section{Introduction}

%



It is  known that full-duplex (FD) communication has a potential to  double spectral efficiency  via simultaneous transmission and reception (STR) over the same frequency band. With recent success in developing FD communication prototypes~\cite{Jain,Duarte,Katti}, there is a renewed  interest in FD studies~\cite{RFeng2016,ZhouY,YWang2015,SP2015,QLi2016,FZhu2014,YSun2015,Gzheng2013,WangY}. Among these, exploiting full duplexity to enhance physical-layer (PHY) security has received considerable attention.

%

PHY security is an information theoretical approach to achieving confidentiality at the PHY \cite{YLiang2008}. A key performance measure of PHY security is the {\it  secrecy rate} --- at which the confidential information can be securely and successfully transmitted from the source to the legitimate destination. By exploiting full duplexity at the transceiver, it will not only help improve  the  secrecy rate, but also provides more flexibility for system designs.
To be specific, the work \cite{Gzheng2013} studied a point-to-point secure communication, where an FD target user  simultaneously receives information from the transmitter and sends artificial noise (AN) to block the eavesdropper. It is shown that higher secrecy rate can be achieved as compared with the conventional transmitter side jamming strategy, owing to a priori knowledge of AN or self interference (SI) at the receiver. Following this idea, the work~\cite{ZhouY} further studied antenna configurations for information reception and AN generation at the FD receiver.
Besides the one-way transmission, two-way secrecy designs were considered in~\cite{RFeng2016,YWang2015} under both perfect and imperfect channel state information (CSI) of the eavesdropper. A semidefinite programming (SDP)-based search approach was proposed to maximize the two-way sum secrecy rate (SSR). The result shows that for well suppressed SI, FD has a better secrecy performance than half duplex. Recently, FD relay secure communications also gained much interest. In~\cite{SP2015}, the authors  considered an adaptive FD one-way relay network, where the relay can  work under either the FD transmission (FDT) mode or the FD jamming (FDJ) mode, depending on the residual SI level and the relative channel quality between the target user and the  eavesdropper. In our recent work \cite{QLi2016}, a joint FDT and FDJ relaying scheme based on Alamouti rank-two beamforming was proposed to enhance PHY security for an FD two-way relay network. Full duplexity  also provides flexibility for secrecy design in cellular networks. By assuming full duplexity at the base station (BS), the uplink (respectively downlink) transmission can be   protected from eavesdropping by deliberately sending AN in the downlink (respectively uplink) transmission. This idea was applied in~\cite{FZhu2014} and \cite{YSun2015} for uplink-downlink SSR maximization and  power minimization respectively (resp.), and more recently was generalized to the simultaneous wireless information and power transfer (SWIPT) scenario~\cite{WangY}.

In this work, we focus on the FD bidirectional secure communications, where two FD legitimate nodes, named Alice and Bob, exchange confidential information, and an eavesdropper, named Eve, overhears the transmissions. Assuming that Alice and Bob both have $N$ transmit antennas and one receive antenna, we aim to optimize the transmit covariance matrices at Alice and Bob so that the sum secrecy rate (SSR) of the bidirectional transmissions is maximized.
This sum secrecy rate maximization (SSRM) problem is nonconvex and involves two matrix variables of dimension $N$-by-$N$. By carefully examining the SSRM problem structure, we first show that the variables' dimension can be reduced from $N$-by-$N$ to 3-by-3. Based on this dimension-reduced formulation, an alternating difference-of-concave (ADC) programming approach is proposed to iteratively optimize the transmit covariance matrices. In particular, we custom-derive a semi-closed-form solution for each ADC iteration and  show that the ADC approach is guaranteed to converge to a stationary solution of the SSRM problem.


Besides the SSRM problem, we also study a robust SSRM problem by assuming imperfect CSI of Eve. In the existing robust secrecy studies, there are  two popular robust models, namely, the bounded-sphere model~\cite{RFeng2016,qli11} and the Gaussian random model~\cite{Rome,Jorswieck,qli14}. In this paper, we depart from the aforementioned models and consider another \emph{moment-based} random CSI model; that is, only some roughly estimated first and second-order statistics of Eve's CSI are available, but the exact distribution and other high-order statistics are not known. Such a model is motivated by the observation that it is relatively easier to estimate the mean and covariance than the complete distribution. Under this moment-based CSI model, we aim to maximize the SSR while keeping  the secrecy outage probability, evaluated with respect to (w.r.t.) any distribution fulfilling the estimated first and second-order statistics, below a given threshold. The considered robust SSRM formulation has two distinguishing
features: 1) It does not require full knowledge of Eve's CSI distribution, thus bypassing the troublesome
distribution modeling problem. 2) it renders a secrecy design that is immune to variations of the distribution,
thus providing a distributionally robust secrecy outage probability guarantee. The latter is particularly important for
PHY security, where the information leakage should be stringently controlled in a worst-case sense. Despite its attractive features, the robust SSRM problem is, however, challenging to solve, because the outage probability generally has no closed form; even
if it has, the resulting constraint is likely to be non-convex. Moreover, the outage probability should
be satisfied for infinite-many distributions with given first and second-order statistics, which thus gives rise to an
infinite number of probabilistic constraints. To handle these difficulties, we employ the Lagrangian duality theory~\cite{Ye} (see also~\cite{YCWu}) and the DC programming to derive a tractable safe solution for the robust SSRM problem.

\subsection{Related Works}
There are some related works worth mentioning. In~\cite{RFeng2016,YWang2015}, the authors  considered similar secrecy design problems under both perfect and imperfect CSI of Eve. Our work differs from~\cite{RFeng2016,YWang2015} in both problem formulation and solution approach. Specifically, for the perfect CSI case the work~\cite{YWang2015} considered the SSRM problem under a single total power constraint, and proposed an SDP-based two-dimensional search approach and some
low-complexity solutions to  the SSRM problem. Herein, we consider the SSRM under the individual power constraints on Alice and Bob. Because of the individual power constraints, the approach in~\cite{YWang2015} is not applicable.
For the imperfect CSI case, our moment-based random CSI model is different from the  bounded-sphere model in \cite{RFeng2016}. Consequently, a completely different robust SSRM formulation as well as the solution method are sought.

\subsection{Organization and Notations}
This paper is organized as follows. The system model and problem statement are given in Section~\ref{sec:model}.
Section~\ref{sec:SSRM} focuses on the SSRM problem under the perfect CSI case and develops an alternating DC approach.
Section~\ref{sec:rob_ssrm} studies the robust SSRM problem under the imperfect CSI case, and develops a robust DC approach.
Simulation results comparing the proposed designs are illustrated in Section~\ref{sec:sim}. Section~\ref{sec:conclusion} concludes the paper.

Our notations are as follows.
$(\cdot)^T$ and $(\cdot)^H$ denote the transpose and conjugate transpose, resp.;  $\mathbf{I}$ denotes an identity matrix with an appropriate dimension; $\bm 1_N$ represents a length-$N$ vector with each entry being one; $\mathbb{H}_{+}^{N}$ (resp. $\mathbb{H}_{++}^{N}$) denotes the set of all $N$-by-$N$ Hermitian positive semidefinite (resp. positive definite) matrices; $\mathbf{A}\succeq \mathbf{0}$ means that $\mathbf{A}$ is Hermitian positive semidefinite, and ${\bf A} \preceq {\bf 0}$ means that $-\mathbf{A}$ is Hermitian positive semidefinite;
${\rm Diag}(\bm A, ~\bm B)$ represents a block diagonal matrix with the diagonal blocks $\bm A$ and $\bm B$; ${\rm Tr}(\cdot)$ denotes a trace operation; $[\cdot]^+ \triangleq \max\{0,\cdot\}$;
$\mathcal{CN}(\bm a, \bm \Sigma)$ represents a complex Gaussian distribution with mean $\bm a$ and covariance matrix $\bm \Sigma$.

\section{System Model and Problem Statement} \label{sec:model}

\begin{figure}[!htp]
  \centering
  \includegraphics[width=2.6in]{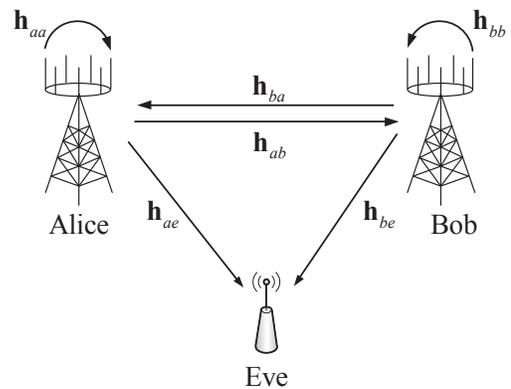}
  \caption{System model}\label{fig:model}
  \vspace{-5pt}
\end{figure}

Consider a full-duplex bidirectional secure communication system depicted in Fig.~\ref{fig:model}, where two legitimate nodes, named Alice and Bob, exchange confidential information, and an eavesdropper, named Eve, overhears the transmissions. We assume that Alice and Bob both work in the full-duplex mode, i.e., Alice (resp. Bob) can simultaneously transmit information to and receive from Bob (resp. Alice). Alice and Bob both have $N$ transmit antennas  and one receive antenna. Eve has a single antenna; extension to multiple multiantenna Eves will be considered in Section~\ref{sec:extension}.
Let $\bm h_{ab} \in \mathbb{C}^N$, $\bm h_{ae}\in \mathbb{C}^N$ and $\bm h_{aa}\in \mathbb{C}^N$ be channels  from Alice  to  Bob, Eve and Alice herself, resp.; $\bm h_{ba} \in \mathbb{C}^N$, $\bm h_{be}\in \mathbb{C}^N$ and $\bm h_{bb}\in \mathbb{C}^N$ are defined similarly. Then, the received signal at Alice can be expressed as
\begin{equation} \label{eq:rx_model}
    y_a(t) = \bm h_{ba}^H \bm x_b(t) + \bm h_{aa}^H \bm x_a(t) + n_a(t),
\end{equation}
where $n_a(t)\sim \mathcal{CN}(0,\sigma_a^2)$ is circularly symmetric complex Gaussian noise at Alice; $\bm x_a(t)\in \mathbb{C}^N$ and $\bm x_b(t) \in \mathbb{C}^N$ are coded confidential information sent by Alice and Bob, resp. According to \cite{YLiang2008}, we have $\bm x_a(t) \sim \mathcal{CN}(\bm 0, \bm Q_a)$ and $\bm x_b(t) \sim \mathcal{CN}(\bm 0, \bm Q_b)$ with $\bm Q_a\succeq \bm 0$ and $\bm Q_b\succeq \bm 0$ being the transmit covariance matrices. The second term on the right-hand side of~\eqref{eq:rx_model} is SI induced by the full-duplex operation. Ideally, the SI can be completely eliminated by exploiting a priori knowledge of $\bm x_a(t)$ at Alice. However, in practice, it can  be only suppressed to some extent, owing to high SI power and hardware limitations, e.g., limited dynamic range. As such, the received signal-to-interference-plus-noise ratio (SINR) at Alice after SI suppression may be expressed as
\begin{equation*}
  {\sf SINR}_a(\bm Q_a, \bm Q_b)= \frac{\bm h_{ba}^H \bm Q_b \bm h_{ba}}{\sigma_a^2 + \zeta_a \bm h_{aa}^H \bm Q_a \bm h_{aa}}
\end{equation*}
where $0<\zeta_a <1$ denotes the SI residual factor, which reflects the residual SI power level after SI suppression. Accordingly, the achievable rate at Alice is given by
\begin{equation} \label{eq:def_Ra}
R_a(\bm Q_a, \bm Q_b)= \log (1 + {\sf SINR}_a(\bm Q_a, \bm Q_b) ),
\end{equation}
where we have treated the residual SI as Gaussian noise.

Similarly, the SINR  at Bob is given by
\begin{equation*}
  {\sf SINR}_b(\bm Q_a, \bm Q_b)= \frac{\bm h_{ab}^H \bm Q_a \bm h_{ab}}{\sigma_b^2 + \zeta_b \bm h_{bb}^H \bm Q_b \bm h_{bb}},
\end{equation*}
and the achievable rate is
\begin{equation}\label{eq:def_Rb}
  R_b(\bm Q_a, \bm Q_b)= \log (1 + {\sf SINR}_b(\bm Q_a, \bm Q_b) ).
\end{equation}
As for Eve, by using the two-user multiple access channel capacity result~\cite{Tse}, its sum rate can be upper bounded by
\begin{equation}\label{eq:def_Re}
R_e(\bm Q_a, \bm Q_b) = \log \left( 1 + \frac{\bm h_{ae}^H \bm Q_a \bm h_{ae} + \bm h_{be}^H \bm Q_b \bm h_{be} }{\sigma_e^2}\right).
\end{equation}
Hence, according to~\cite{Yener} the sum secrecy rate of the bidirectional transmissions is expressed as
\[ R_s(\bm Q_a, \bm Q_b) =  [R_a(\bm Q_a, \bm Q_b) +  R_b(\bm Q_a, \bm Q_b)-R_e(\bm Q_a, \bm Q_b)]^+ .\]

Our problem of interest is to jointly optimize  $\bm Q_a$ and $\bm Q_b$ such that the sum secrecy rate  is maximized, viz.,
\begin{subequations}\label{eq:ssrm}
  \begin{align}
   \max_{\bm Q_a, \bm Q_b} & ~ R_s(\bm Q_a, \bm Q_b)  \label{eq:ssrm-a} \\
\hspace{-10pt}   {\sf (SSRM)}~~ {\rm s.t.} & ~ {\rm Tr}(\bm Q_a) \leq P_a, ~~\bm Q_a \succeq \bm 0, \label{eq:ssrm-b}\\
    & ~ {\rm Tr}(\bm Q_b) \leq P_b, ~~\bm Q_b \succeq \bm 0, \label{eq:ssrm-c}
  \end{align}
\end{subequations}
where $P_a>0$ and $P_b>0$ denote  the total transmit power budget at Alice and Bob, resp.


\section{A Tractable Approach to Problem~\eqref{eq:ssrm}} \label{sec:SSRM}

\subsection{Reformulation of Problem~\eqref{eq:ssrm}}\label{sec:sol_chara}
In~\eqref{eq:ssrm}, the variables' dimension is $N$-by-$N$. By exploiting the problem structure, problem~\eqref{eq:ssrm} can be recast into a form with the variables' dimension of $3$-by-$3$. Specifically, from the objective~\eqref{eq:ssrm-a}, it is not hard to see that the optimal $\bm Q_a$ must lie in the  subspace spanned by $\bm H_a \triangleq [\bm h_{ab},~ \bm h_{aa},~ \bm h_{ae}]$. This is because $R_a$, $R_b$ and $R_e$ depend on $\bm Q_a$ through $\bm h_{aa}^H \bm Q_a \bm h_{aa}$, $\bm h_{ab}^H \bm Q_a \bm h_{ab}$ and $\bm h_{ae}^H \bm Q_a \bm h_{ae}$, resp.
If $\bm Q_a$ has any component outside the range space of $\bm H_a$, we can project the former onto the latter without changing the objective value in \eqref{eq:ssrm-a}.
Therefore, without loss of optimality, we can set
\begin{equation} \label{eq:Qa_structure}
  \bm Q_a = \bm U_a \bm W_a \bm U_a^H
\end{equation}
for some $\bm W_a \in \mathbb{H}_+^3$. Here, $\bm U_a \in \mathbb{C}^{N\times 3}$ is any semi-unitary matrix spanning the same subspace as $\bm H_a$, which can be obtained by standard Gram-Schmidt process. Similarly, the optimal $\bm Q_b$ is given by
\begin{equation} \label{eq:Qb_structure}
  \bm Q_b = \bm U_b \bm W_b \bm U_b^H
\end{equation}
for some $\bm W_b \in \mathbb{H}_+^3$, where $\bm U_b\in \mathbb{C}^{N\times3}$ spans the same range space as $\bm H_b \triangleq [\bm h_{ba},~ \bm h_{bb},~ \bm h_{be}]$.

Using~\eqref{eq:Qa_structure} and \eqref{eq:Qb_structure}, problem~\eqref{eq:ssrm} can be equivalently expressed as
\begin{equation}\label{eq:ssrm_eqv}
\begin{aligned}
   \max_{\bm W_a \in \mathbb{H}^3, \bm W_b \in \mathbb{H}^3} & ~ \tilde{R}_a(\bm W_a, \bm W_b) + \tilde{R}_b(\bm W_a, \bm W_b) - \tilde{R}_e(\bm W_a, \bm W_b)  \\
    {\rm s.t.} &  ~ {\rm Tr}(\bm W_a) \leq P_a, ~~ \bm W_a \succeq\bm 0, \\
    & ~ {\rm Tr}(\bm W_b) \leq P_b, ~~\bm W_a \succeq\bm 0,
    \end{aligned}
\end{equation}
where
\begin{align*}
  \tilde{R}_a(\bm W_a, \bm W_b) &  = \log\left( 1 + \frac{\tilde{\bm h}_{ba}^H \bm W_b \tilde{\bm h}_{ba}}{\sigma_a^2 + \zeta_a \tilde{\bm h}_{aa}^H \bm W_a \tilde{\bm h}_{aa}} \right), \\
  \tilde{R}_b(\bm W_a, \bm W_b) &  = \log\left( 1 + \frac{\tilde{\bm h}_{ab}^H \bm W_a \tilde{\bm h}_{ab}}{\sigma_b^2 + \zeta_b \tilde{\bm h}_{bb}^H \bm W_b \tilde{\bm h}_{bb}} \right), \\
  \tilde{R}_e(\bm W_a, \bm W_b) & = \log \left( 1 + \frac{\tilde{\bm h}_{ae}^H \bm W_a \tilde{\bm h}_{ae} + \tilde{\bm h}_{be}^H \bm W_b \tilde{\bm h}_{be} }{\sigma_e^2}\right),
\end{align*}
and $\tilde{\bm h}_{ij} \triangleq \bm U_i^H \bm h_{ij} \in \mathbb{C}^{3}, \forall~i\in \{a, b\}, j\in \{a,b,e\}$.

By comparing \eqref{eq:ssrm_eqv} with \eqref{eq:ssrm}, the number of variables has been   reduced to 18 real variables. However, computing these optimal 18 variables  is still a challenging task, owing to the highly nonlinear objective.  In the sequel, we propose  a suboptimal, yet computationally efficient approach to problem~\eqref{eq:ssrm_eqv} by using alternating optimization and the DC programming.

\subsection{An Alternating DC Approach to Problem~\eqref{eq:ssrm_eqv}}
Since it is hard to jointly  optimize $\bm W_a$ and $\bm W_b$, a natural idea is to alternately  optimizing $\bm W_a$ and $\bm W_b$. Suppose that at the $k$th iteration we have obtained $(\bm W_a^k, \bm W_b^k)$. Then, we alternately solve the following two problems
\begin{equation} \label{eq:AO_a}
  \begin{aligned}
    \bm W_a^{k+1} \in  \arg\max_{\bm W_a }  & ~  \Big\{ \tilde{R}_a(\bm W_a, \bm W_b^k) + \tilde{R}_b(\bm W_a, \bm W_b^k) \\
    & ~~~ - \tilde{R}_e(\bm W_a, \bm W_b^k) \Big\}\\
     {\rm s.t.}    &~ {\rm Tr}(\bm W_a) \leq P_a,~~\bm W_a \succeq \bm 0,
  \end{aligned}
\end{equation}
and
\begin{equation}\label{eq:AO_b}
  \begin{aligned}
    \bm W_b^{k+1} \in  \arg\max_{\bm W_b} & ~  \Big\{ \tilde{R}_a(\bm W_a^{k+1}, \bm W_b) + \tilde{R}_b(\bm W_a^{k+1}, \bm W_b)  \\
    & ~~~ - \tilde{R}_e(\bm W_a^{k+1}, \bm W_b) \Big\}\\
    {\rm s.t.}  &  ~ {\rm Tr}(\bm W_b) \leq P_b,~~\bm W_b\succeq \bm 0,
  \end{aligned}
\end{equation}
for $k=1,2,\ldots$ until some stopping criterion is satisfied. By directly applying the  convergence result~\cite{Grippo} of  block-coordinate descent (BCD) with two blocks, the following fact is readily obtained.
\begin{Fact}[\hspace{-0.5pt}\cite{Grippo}]\label{fact}
  Suppose that $\{(\bm W_a^k, \bm W_b^k)\}_k$ is a sequence generated by alternately solving \eqref{eq:AO_a} and \eqref{eq:AO_b}. Then, every limit point of $\{(\bm W_a^k, \bm W_b^k)\}_k$ is a stationary point of problem~\eqref{eq:ssrm_eqv}.
\end{Fact}
We should mention that  the above alternating optimization  requires solving each subproblem in \eqref{eq:AO_a} and \eqref{eq:AO_b}  optimally. However, one can verify that the subproblems in~\eqref{eq:AO_a} and \eqref{eq:AO_b} are still nonconvex. In particular, the objective in \eqref{eq:AO_a} is a difference of concave (DC) functions, whereby $\tilde{R}_b(\bm W_a, \bm W_b^k) $ is concave w.r.t. $\bm W_a$, but $\tilde{R}_a(\bm W_a, \bm W_b^k) - \tilde{R}_e(\bm W_a, \bm W_b^k)$ is convex. A similar observation applies to \eqref{eq:AO_b}. A widely-used approach to handle this kind of problem is the DC programming, i.e., by locally linearizing the nonconcave part $\tilde{R}_a(\bm W_a, \bm W_b^k) - \tilde{R}_e(\bm W_a, \bm W_b^k)$ to get a convex approximation of problem~\eqref{eq:AO_a}. This motivates us to consider an alternating DC (ADC) approach to problem~\eqref{eq:ssrm_eqv}; see Algorithm~\ref{algorithm:1}.
\begin{algorithm}
  \caption{An ADC Approach to Problem~\eqref{eq:ssrm_eqv}}\label{algorithm:1}
\begin{algorithmic}[1]
  \State Set $k=0$ and initialize $(\bm W_a^{0}, \bm W_b^{0})$
   \Repeat
  \begin{equation} \label{eq:dc_a}
  \begin{aligned}
   \bm W_a^{k+1} \in \arg \max_{\bm W_a \succeq \bm 0}  & ~ f_a(\bm W_a; \bm W_a^k,\bm W_b^k) \\
     {\rm s.t.}    &~ {\rm Tr}(\bm W_a) \leq P_a,
  \end{aligned}
\end{equation}
\State \begin{equation}\label{eq:dc_b}
  \begin{aligned}
    \bm W_b^{k+1} \in \arg \max_{\bm W_b\succeq \bm 0} & ~  f_b(\bm W_b; \bm W_a^{k+1},\bm W_b^k)  \\
    {\rm s.t.}  &  ~ {\rm Tr}(\bm W_b) \leq P_b,
  \end{aligned}
\end{equation}
\State $k=k+1$;
\Until{Some stopping criterion is satisfied}
\end{algorithmic}
\end{algorithm}
Note that in~\eqref{eq:dc_a} and \eqref{eq:dc_b}, we have defined
 \begin{align*}
   & f_a(\bm W_a; \bm W_a^k,\bm W_b^k) \\
   \triangleq & \tilde{R}_b(\bm W_a, \bm W_b^k)  +  \tilde{R}_a(\bm W_a^k, \bm W_b^k) - \tilde{R}_e(\bm W_a^k, \bm W_b^k) \\
  +  &  {\rm Tr}\left(\nabla_{\bm W_a} [\tilde{R}_a(\bm W_a^k, \bm W_b^k) - \tilde{R}_e(\bm W_a^k, \bm W_b^k) ]^H  (\bm W_a- \bm W_a^k) \right)\\
   & f_b(\bm W_b; \bm W_a^{k+1},\bm W_b^k) \\
   \triangleq & \tilde{R}_a(\bm W_a^{k+1}, \bm W_b)  + \tilde{R}_b(\bm W_a^{k+1}, \bm W_b^k) - \tilde{R}_e(\bm W_a^{k+1}, \bm W_b^k) \\
   + &  {\rm Tr} \left(\nabla_{\bm W_b} [\tilde{R}_b(\bm W_a^{k+1}, \bm W_b^k) - \tilde{R}_e(\bm W_a^{k+1}, \bm W_b^k) ]^H (\bm W_b- \bm W_b^k) \right).
 \end{align*}

In steps 2 and 3 of Algorithm~\ref{algorithm:1}, we need to solve two convex optimization problems, which in principle can be done by invoking some general purpose optimization softwares, such as~\texttt{CVX}~\cite{cvx}. However, by carefully inspecting the problem structure, we are able to custom-derive  more efficient solutions for problems \eqref{eq:AO_a} and  \eqref{eq:AO_b}, as detailed in the next subsection.


\subsection{Semi-closed-form Solutions for Problems~\eqref{eq:dc_a} and \eqref{eq:dc_b}}\label{sec:fast_solution}
Let us focus on problem~\eqref{eq:dc_a} and the other one can be handled similarly. After dropping some terms irrespective of $\bm W_a$, problem~\eqref{eq:dc_a} is simplified as
  \begin{equation} \label{eq:dc_a_eqv}
  \begin{aligned}
    \max_{\bm W_a \succeq \bm 0}  & ~ \log(1 + \hat{\bm h}_{ab}^H \bm W_a \hat{\bm h}_{ab} ) - {\rm Tr}(\bm M_a \bm W_a) \\
     {\rm s.t.}    &~ {\rm Tr}(\bm W_a) \leq P_a,
  \end{aligned}
\end{equation}
where $\hat{\bm h}_{ab}$ and $\bm M_a$ are defined in~\eqref{eq:Ma_def}. Let us consider solving the dual of problem~\eqref{eq:dc_a_eqv}, which is given by
\begin{figure*}[!t]
\setlength\arraycolsep{0pt}
\begin{equation}\label{eq:Ma_def}
  \begin{aligned}
 & \hat{\bm h}_{ab} = ({\sigma_b^2 + \zeta_b \tilde{\bm h}_{bb}^H \bm W_b^{k} \tilde{\bm h}_{bb}})^{-1/2} \tilde{\bm h}_{ab}, ~ \hat{\bm h}_{aa} = \sqrt{{\zeta_a}/ (\tilde{\bm h}_{ba}^H \bm W_b^{k} \tilde{\bm h}_{ba})} \tilde{\bm h}_{aa}, ~ \hat{\bm h}_{ae} = \tilde{\bm h}_{ae}/\sqrt{\sigma_e^2 + \tilde{\bm h}_{be}^H \bm W_b^k \tilde{\bm h}_{be} }, ~ \hat{\sigma}_a^2 = \sigma_a^2/(\tilde{\bm h}_{ba}^H \bm W_b^k \tilde{\bm h}_{ba}) \\
 & \hspace{3cm}\bm M_a = \frac{\hat{\bm h}_{aa}\hat{\bm h}_{aa}^H }{(1 + \hat{\sigma}_a^2 + \hat{\bm h}_{aa}^H \bm W_a^k \hat{\bm h}_{aa}) (\hat{\sigma}_a^2 + \hat{\bm h}_{aa}^H \bm W_a^k \hat{\bm h}_{aa})} + \frac{\hat{\bm h}_{ae} \hat{\bm h}_{ae}^H}{1 + \hat{\bm h}_{ae}^H \bm W_a^k \hat{\bm h}_{ae}}
\end{aligned}
\end{equation}
 \hrulefill
\end{figure*}
\begin{equation}\label{eq:dual}
  \min_{\lambda_a \geq 0} \left\{ \max_{\bm W_a\succeq \bm 0}  ~ \mathcal{L}(\bm W_a, \lambda_a) \right\},
\end{equation}
where
$\mathcal{L}(\bm W_a, \lambda_a)  \triangleq \log(1 + \hat{\bm h}_{ab}^H \bm W_a \hat{\bm h}_{ab} ) - {\rm Tr}(\bm M_a \bm W_a) - \lambda_a ({\rm Tr}(\bm W_a)- P_a) $
is the partial Lagrangian of problem~\eqref{eq:dc_a_eqv}, and  $\lambda_a\geq 0$ is the Lagrangian multiplier associated with the power constraint.
Depending on the relationship between $\hat{\bm h}_{ab}$ and $\bm M_a$, we consider the solution of \eqref{eq:dual} in the following two cases, namely, $\hat{\bm h}_{ab} \notin {\cal R}\{\bm M_a\}$ and  $\hat{\bm h}_{ab} \in {\cal R}\{\bm M_a\}$, where ${\cal R}\{\bm M_a\}$ denotes the range space spanned by $\bm M_a$.

\subsubsection{Case 1: $\hat{\bm h}_{ab} \notin {\cal R}\{\bm M_a\}$} In such a case, any dual feasible $\lambda_a$ of \eqref{eq:dual} must be strictly positive, because for $\lambda_a=0$, we can always find some $\bm v$ such that  $\bm M_a \bm v = \bm 0$ and $\hat{\bm h}_{ab}^H \bm v \neq 0$. It is easy to verify that ${\cal L}(\zeta \bm v \bm v^H, 0) \rightarrow \infty$  as $\zeta \rightarrow \infty$; that is $\lambda_a=0$ is dual infeasible for~\eqref{eq:dual}. In view of this, we charaterize the primal-dual optimal solution of \eqref{eq:dual} as follows.
\begin{Prop}\label{lemma:closed_form_sol}
 Suppose $\hat{\bm h}_{ab} \notin {\cal R}\{\bm M_a\}$. The optimal  $\bm W_a^\star$ of \eqref{eq:dc_a_eqv} is given by
   \begin{equation}\label{eq:closed_form_sol_Wa}
 \bm W_a^\star = \kappa_a (\bm M_a+ \lambda_a^\star \bm I)^{-1}  \hat{\bm h}_{ab} \hat{\bm h}_{ab}^H  (\bm M_a+ \lambda_a^\star \bm I)^{-1},
\end{equation}
where $\kappa_a =  \frac{{\big[1- {\|  (\bm M_a+ \lambda_a^\star \bm I)^{-1/2} \hat{\bm h}_{ab} \|^{-2}} \big]^+}}{{ \| (\bm M_a+ \lambda_a^\star \bm I)^{-1/2} \hat{\bm h}_{ab}\|^2}}$ and $\lambda_a^\star$ is the dual optimal solution satisfying $0<\lambda_a^\star < \| \hat{\bm h}_{ab}\|^2$. Moreover,  $\lambda_a^\star$ can be efficiently computed via bisection such that ${\rm Tr}(\bm W_a^\star) = P_a $ holds.
 \end{Prop}
 \noindent{\it Proof.}~See Appendix~\ref{appendix:closed_form_sol}. \hfill $\blacksquare$

Some remarks on Proposition~\ref{lemma:closed_form_sol} are in order:
\begin{Remark}\label{remark:2}
 Under the assumption of $\hat{\bm h}_{ab} \notin {\cal R}\{\bm M_a\}$, the transmitter must transmit with full power. Intuitively, this is reasonable because if there is remaining power, one can  utilize it and send information in the null space of $\bm M_a$, which may further improve the objective of \eqref{eq:dc_a_eqv}.
\end{Remark}

\begin{Remark}\label{remark:1}
The optimal $\bm W_a^\star$ has a  rank-one structure, and thus the ADC  algorithm can be implemented with the simple transmit beamforming. Moreover, the beamformer $\sqrt{\kappa_a} (\bm M_a+ \lambda_a^\star \bm I)^{-1}  \hat{\bm h}_{ab}$ has an interesting connection with the MMSE-based  transmit beamforming in interfering channels (IC)~\cite{Wangxd}. In particular, if we treat  $\hat{\bm h}_{ab}$ as the target channel from the transmitter to Bob, and $\hat{\bm h}_{aa}$ and $\hat{\bm h}_{ae}$ as interfering channels to Alice herself and Eve, resp. Then, $\bm M_a$ is exactly the sum of the covariance matrices  of interferences seen at Alice and Eve. Therefore, the beamforming solution $\sqrt{\kappa_a} (\bm M_a+ \lambda_a^\star \bm I)^{-1}  \hat{\bm h}_{ab}$ trades off the target user's reception and  ``interference'' suppression at (or information leakage to) co-channel users. Specifically, it can be easily shown that when the co-channels are absent, i.e., $\hat{\bm h}_{aa}=\hat{\bm h}_{ae}=\bm 0$, the beamformer degenerates into the maximum ratio transmission (MRT). On the other hand, when the intensities of the co-channels are extremely strong, the optimal beamforming approaches  zero-forcing beamforming.
\end{Remark}

\begin{Remark}\label{remark:3}
The rank-one structure of $\bm W_a^\star$ is not only physically meaningful, it also plays a key role in  pining down the convergence of the ADC algorithm. We will detail this later in Proposition~\ref{prop:adc_convergence}.
\end{Remark}

\subsubsection{Case 2: $\hat{\bm h}_{ab} \in {\cal R}\{\bm M_a\}$} In such a case, the dual feasible solution $\lambda_a$ is not necessarily strictly positive. Nevertheless, since $\hat{\bm h}_{ab} \in {\cal R}\{\bm M_a\}$ and the objective of \eqref{eq:dc_a_eqv} depends on $\bm W_a$  through $\hat{\bm h}_{ab}$ and $\bm M_a$, without loss of optimality, we may assume that the optimal $\bm W_a^\star$ lies in the range space of $\bm M_a$, for otherwise we can project $\bm W_a^\star$ onto the range space of $\bm M_a$ and attains the same optimal value with less power. Let $\bm M_a = \bm F_a \bm \Sigma_a \bm F_a^H$ be the economy SVD of $\bm M_a$ with $\bm F_a \in \mathbb{C}^{3\times 2}$ and $\bm \Sigma_a \in \mathbb{H}_{++}^2$. Then, the optimal $\bm W_a^\star$ takes the following form:
 \begin{equation} \label{eq:W__X_relation}
   \bm W_a^\star = \bm F_a \bm X_a^\star \bm F_a^H
 \end{equation}
for some $\bm X_a^\star \in \mathbb{H}_+^2 $. Using \eqref{eq:W__X_relation}, problem~\eqref{eq:dc_a_eqv} is simplified as
\begin{equation}\label{eq:dc_a_eqv_Xa}
    \begin{aligned}
    \max_{\bm X_a \in \mathbb{H}_+^2 }  & ~ \log(1 + \hat{\bm h}_{ab}^H \bm F_a \bm X_a \bm F_a^H \hat{\bm h}_{ab} ) - {\rm Tr}(\bm \Sigma_a \bm X_a) \\
     {\rm s.t.}    &~ {\rm Tr}(\bm X_a) \leq P_a.
  \end{aligned}
\end{equation}
By noting that $\bm \Sigma_a\succ\bm 0$, a similar result like Proposition~\ref{lemma:closed_form_sol} can be easily deduced.
\begin{Prop}\label{lemma:closed_form_sol_C2}
 Suppose $\hat{\bm h}_{ab} \in {\cal R}\{\bm M_a\}$. The optimal  $\bm W_a^\star$ of \eqref{eq:dc_a_eqv} is given by \eqref{eq:W__X_relation} with $\bm X_a^\star$ computed as
   \begin{equation}\label{eq:closed_form_sol_Xa}
 \bm X_a^\star = \tilde{\kappa}_a (\bm \Sigma_a+ \lambda_a^\star \bm I)^{-1}  \bm F_a^H \hat{\bm h}_{ab} \hat{\bm h}_{ab}^H \bm F_a  (\bm \Sigma_a+ \lambda_a^\star \bm I)^{-1},
\end{equation}
where $\tilde{\kappa}_a =  \frac{{\big[1- {\|  (\bm \Sigma_a+ \lambda_a^\star \bm I)^{-1/2}  \bm F_a^H \hat{\bm h}_{ab} \|^{-2}} \big]^+}}{{ \| (\bm \Sigma_a+ \lambda_a^\star \bm I)^{-1/2} \bm F_a^H \hat{\bm h}_{ab}\|^2}}$ and $\lambda_a^\star \geq 0$ is the optimal dual variable associated with the power constraint in \eqref{eq:dc_a_eqv_Xa}. Moreover,  $\lambda_a^\star$ can be efficiently computed via bisection such that the complementarity condition $\lambda_a^\star ({\rm Tr}(\bm X_a^\star)- P_a)=0 $ is satisfied.
 \end{Prop}

\noindent{\it Proof.}~Notice that $\bm \Sigma_a\succ\bm 0$ implies that $\bm \Sigma_a+ \lambda_a^\star \bm I$ is invertible, whenever $\lambda_a^\star \geq 0$. Then, the remaining proof is exactly the same as Proposition~\ref{lemma:closed_form_sol}, and thus omitted for brevity. \hfill $\blacksquare$

Let us examine the convergence issue of the ADC algorithm. The ADC algorithm is somehow a mixture of the BCD method and the DC programming, but neither of their  convergence results applies. By linking the ADC algorithm to the block successive upper-bound minimization (BSUM) algorithm in \cite{Luo}, the following convergence result of the ADC algorithm can be readily established.
\begin{Prop}\label{prop:adc_convergence}
  The sequence $\{(\bm W_a^k, \bm W_b^k) \}_k$ generated by Algorithm~\ref{algorithm:1} must have at least one limit point. Moreover, every limit point of $\{(\bm W_a^k, \bm W_b^k) \}_k$ is a stationary solution of problem~\eqref{eq:ssrm_eqv}.
\end{Prop}
\noindent {\it Proof.}~See Appendix~\ref{appendix:adc_convergence}. \hfill $\blacksquare$

\subsection{Extensions}\label{sec:extension}
The proposed ADC algorithm can be adapted to more complex eavesdropping scenarios. Below, we showcase some possible extensions.

\subsubsection{Multiple Alice-Bob pairs} Consider interfering channels with $K$ pairs of Alice-Bob links.  Let $\bm h_{a_i,b_j}$ be the channel from the $i$th Alice to the $j$th Bob, and $\bm Q_{a_i}$ be the transmit covariance of the $i$th Alice. $\bm h_{a_i,e}$, $\bm h_{b_i,a_j}$ and $\bm Q_{b_i}$ are defined similarly. The rate of the $i$th Alice is
\begin{align*}
  & R_{a_i}(\{\bm Q_{a_k}, \bm Q_{b_k}\}_{k=1}^K)= \\
  & \log\Big(1+\frac{\bm h_{b_i,a_i}^H \bm Q_{b_i} \bm h_{b_i,a_i}}{\sigma_{a_i}^2 + \zeta_{a_i} \bm h_{a_i}^H \bm Q_{a_i} \bm h_{a_i} + \displaystyle  \sum_{ {j \neq i, x\in\{a,b\}}} \bm h_{x_j,a_i}^H \bm Q_{x_j} \bm h_{x_j,a_i} } \Big)
\end{align*}
The $i$th Bob's rate $R_{b_i}(\{\bm Q_{a_k}, \bm Q_{b_k}\}_{k=1}^K)$ can be calculated similarly. The Eve's sum rate is given by
\begin{align*}
  & R_{e}(\{\bm Q_{a_k}, \bm Q_{b_k}\}_{k=1}^K)\\
  = &  \log \Big(1+ \frac{\sum_{i=1}^K \bm h_{a_i,e}^H \bm Q_{a_i}\bm h_{a_i,e}+ \bm h_{b_i,e}^H \bm Q_{b_i}\bm h_{b_i,e} }{\sigma_e^2} \Big).
\end{align*}
Then, the sum secrecy rate is written as $R_s(\{\bm Q_{a_k}, \bm Q_{b_k}\}_{k=1}^K) = [(\sum_{i=1}^K R_{a_i} + R_{b_i} ) - R_{e}]^+$.  Since there are multiple $\bm Q_{a_i}$ and $\bm Q_{b_i}$, it is straightforward to extend the previous two-block ADC algorithm to $2K$-block ADC. In particular, we cyclically optimize $\bm Q_{x_i}$ for $x\in\{a,b\}$ and $i=1,\ldots, K$. When fixing any $2K-1$ variables and optimizing the remaining one,  the resultant secrecy rate maximization problem is similar to problem~\eqref{eq:AO_a}. Hence, the DC programing and the fast solutions in Propositions~\ref{lemma:closed_form_sol} and \ref{lemma:closed_form_sol_C2} can be directly applied. Moreover, the stationary convergence claim is still valid by noting that the BSUM convergence conditions  still hold for $K\geq 2$.

\subsubsection{Multiple multiantenna Eves} The ADC algorithm can be adapted to multiple multiantenna Eves. For simplicity, we still assume one pair of Alice-Bob; extension to multiple pairs is straightforward. Let $\bm H_{ae_i} \in \mathbb{C}^{N\times L_i}$ (resp. $\bm H_{be_i} \in \mathbb{C}^{N\times L_i}$) be the channel from Alice to the $i$th Eve (resp. Bob to the $i$th Eve), with $L_i$ being the number of antennas at Eve $i$. Then, the rate at the $i$th Eve is given  by
\begin{equation*}
  R_{e_i}(\bm Q_a, \bm Q_b) = \log |\bm I + \sigma_{e_i}^{-2} (\bm H_{ae_i}^H \bm Q_a \bm H_{ae_i} + \bm  H_{be_i}^H \bm Q_b \bm H_{be_i})|
\end{equation*}
for $i=1,\ldots, I$. The sum secrecy rate is written as
\begin{equation*}
\begin{aligned}
  & R_s(\bm Q_a, \bm Q_b) \\
  = &  \min_{i=1,\ldots, I} \{ R_a(\bm Q_a, \bm Q_b) + R_b(\bm Q_a, \bm Q_b)- R_{e_i}(\bm Q_a, \bm Q_b)\}.
  \end{aligned}
\end{equation*}
It is easy to verify that under the multiple multiantenna Eve's case, the dimension reduction  and the ADC algorithm is still applicable. In particular,   by using the same notations as in Sec.~\ref{sec:sol_chara}, the DC subproblem~\eqref{eq:dc_a} is modified as:
  \begin{equation} \label{eq:dc_sub_multi_Eve}
  \begin{aligned}
    \max_{\bm W_a \succeq \bm 0}  & ~ \min_{i=1,\ldots, I} \log(1 + \hat{\bm h}_{ab}^H \bm W_a \hat{\bm h}_{ab} ) - {\rm Tr}(\bm M_{i} \bm W_a) \\
     {\rm s.t.}    &~ {\rm Tr}(\bm W_a) \leq P_a,
  \end{aligned}
\end{equation}
where $\bm M_{i}$ is defined in \eqref{eq:Ma_def2}, and $\tilde{\bm H}_{ae_i}$ and $\tilde{\bm H}_{be_i}$ are defined in a similar way as $\tilde{\bm h}_{ae}$ and $\tilde{\bm h}_{be}$ in \eqref{eq:ssrm_eqv}. Problem~\eqref{eq:dc_sub_multi_Eve} is a nonsmooth optimization problem, but can be reformulated as a smooth one, as stated in the following proposition:
\begin{Prop}\label{prop:lemma_multieve_eqv}
Consider the following problem:
  \begin{equation}\label{eq:multieve_eqv_a}
  \begin{aligned}
      \min_{ \bm \gamma }~& g(\bm \gamma), \\
      {\rm s.t.}~ &   \sum_{i=1}^I \gamma_i =1, \gamma_i \geq 0, i=1,\ldots, I,
      \end{aligned}
  \end{equation}
  where
  \begin{equation}\label{eq:multieve_eqv_b}
  \begin{aligned}
    & g(\bm \gamma) \triangleq  \max_{\substack{ \bm W_a \succeq \bm 0, \\  {\rm Tr}(\bm W_a) \leq P_a}} \log(1 + \hat{\bm h}_{ab}^H \bm W_a \hat{\bm h}_{ab} ) - \sum_{i=1}^I \gamma_i {\rm Tr}(\bm M_{i} \bm W_a).
  \end{aligned}
  \end{equation}
Let $\bm W_a^\star(\bm \gamma)$ be the optimal solution of problem~\eqref{eq:multieve_eqv_b} and $\bm \gamma^\star$ be the optimal solution of \eqref{eq:multieve_eqv_a}. Then,
\begin{enumerate}
  \item $g(\bm \gamma)$ is differentiable w.r.t. $\bm \gamma$ and $\nabla g(\bm \gamma) = - [{\rm Tr}(\bm M_i \bm W_a^\star(\bm \gamma)), \ldots, {\rm Tr}(\bm M_I \bm W_a^\star(\bm \gamma))]^T$;
      \item the optimal solution of \eqref{eq:dc_sub_multi_Eve} is given by $\bm W_a^\star(\bm \gamma^\star)$.
\end{enumerate}
\end{Prop}
\noindent{\it Proof.}~See Appendix~\ref{appendix:lemma_multieve_eqv}. \hfill $\blacksquare$

\begin{figure*}[!t]
\setlength\arraycolsep{0pt}
\begin{equation}\label{eq:Ma_def2}
  \begin{aligned}
& \bm M_{i} = \frac{\hat{\bm h}_{aa}\hat{\bm h}_{aa}^H }{(1 + \hat{\sigma}_a^2 + \hat{\bm h}_{aa}^H \bm W_a^k \hat{\bm h}_{aa}) (\hat{\sigma}_a^2 + \hat{\bm h}_{aa}^H \bm W_a^k \hat{\bm h}_{aa})} +   \tilde{\bm H}_{ae_i} (\sigma_{e_i}^{2} \bm I + \tilde{\bm H}_{ae_i}^H \bm W_a^k \tilde{\bm H}_{ae_i} + \tilde{\bm  H}_{be_i}^H \bm W_b^k \tilde{ \bm H}_{be_i} )^{-1}  \tilde{\bm H}_{ae_i}^H
\end{aligned}
\end{equation}
 \hrulefill
\end{figure*}
Since $g(\bm \gamma)$ is differentiable, the
optimal $\bm \gamma^\star$ of \eqref{eq:multieve_eqv_a} can be efficiently computed via the projected gradient descent method~\cite{Bertsekas}, as long as $g(\bm \gamma)$ and $\nabla g(\bm \gamma)$ can be efficiently evaluated. Fortunately, for the considered problem~\eqref{eq:multieve_eqv_a}, $g(\bm \gamma)$ and $\bm W_a^\star(\bm \gamma)$ can be easily computed with the aid of Props.~\ref{lemma:closed_form_sol} and \ref{lemma:closed_form_sol_C2}, if one notices that problem~\eqref{eq:multieve_eqv_b} is exactly the same as problem~\eqref{eq:dc_a_eqv} by replacing $\bm M_a$ with $\sum_{i=1}^I \gamma_i \bm M_i$.  In addition, the projection onto the unit simplex in \eqref{eq:multieve_eqv_a} can also be efficiently performed (in a semi-closed form), say by using the algorithm in~\cite{Duchi08}.

\section{Robust Sum Secrecy Rate Maximization}\label{sec:rob_ssrm}

In the previous SSRM design, we have implicitly assumed perfect knowledge of Eve's CSIs.  However, in practice it may not be easy to acquire the exact CSI. In this section, we  relax this assumption  and consider a robust SSRM design under  imperfect CSI of Eve.

\subsection{ CSI Uncertainty Model and Robust SSRM  Formulation}

As mentioned in Introduction, we focus on  a moment-based random CSI model ---
Eve's CSIs $\bm h_{ae}$ and $\bm h_{be}$  are randomly and independently distributed over the measurable space $\mathbb{C}^N$. We have only some rough estimate of the first and the second-order moment information about $\bm h_{ae}$ and $\bm h_{be}$, but the  exact distribution or other high-order statistics of $\bm h_{ae}$ and $\bm h_{be}$  are not known. A more precise mathematical description of the above moment-based uncertain CSI   model is as follows. Let $\bm \xi_i \in \mathbb{C}^N$ and $\bm \Omega_i \in \mathbb{H}_+^N$ be the estimates of the first and the second-order moments of $\bm h_{ie}$ for $i\in \{a, b\}$, resp. The roughness of these estimates is characterized by the following inequalities:
\begin{subnumcases}{\label{eq:csi_mode_2}}
\| \mathbb{E}[  \bm h_{ie}] -   {\bm \xi}_{i} \|_2 \leq \tau_{i,1},  \label{eq:csi_mode_2_a}\\
\|  \mathbb{E}[   \bm h_{ie}      \bm h_{ie}^H] -  {\bm \Omega}_i \|_2  \leq  \tau_{i,2}, \label{eq:csi_mode_2_b}
  \end{subnumcases}
for $i\in \{a,b\}$. Herein, $\| \cdot\|_2$  denotes either $\ell_2$  norm for vector input or spectral norm for matrix input; $\tau_{i,1} \geq 0$ and $\tau_{i,2} \geq 0$ are given constants, specifying how accurate the  first and the second-order moments are. In particular, $\tau_{i,1}= \tau_{i,2} = 0$ corresponds to ideal estimation.
Notice that given $\bm \xi_i, \bm \Omega_i, \tau_{i,1}$ and $\tau_{i,2}$, there are in general infinite number of distributions  fulfilling \eqref{eq:csi_mode_2}. In other words, the true distribution of $\bm h_{ie}$, denoted by $F_i$ for $i\in \{a,b\}$, lies in an uncertainty set, i.e.,
\begin{equation}\label{eq:def_D}
  F_i \in \mathscr{D} (  {\bm \xi}_i,  {\bm \Omega}_i,  \tau_{i,1}, \tau_{i,2}), \quad i\in \{a,b\},
\end{equation}
where $\mathscr{D} (  {\bm \xi}_i,  {\bm \Omega}_i,  \tau_{i,1}, \tau_{i,2})$ denotes the set of distributions that fulfills \eqref{eq:csi_mode_2}.


Under the above uncertain CSI model, the robust SSRM  problem is formulated as an outage-probability constrained secrecy rate maximization problem, viz.,
\begin{subequations} \label{eq:rrssrm}
\begin{align}
 & \max_{\bm Q_a, \bm Q_b, R_s }  ~  R_s   \label{eq:rrssrm-a} \\
 & {\rm s.t.} ~   \min_{\substack{  F_a \in \mathscr{D} (  {\bm \xi}_a,  {\bm \Omega}_a,  \tau_{a,1}, \tau_{a,2})\\   F_b \in \mathscr{D} (  {\bm \xi}_b,  {\bm \Omega}_b,  \tau_{b,1}, \tau_{b,2})}} \mathbb{P}_{\substack{  \bm h_{ae} \sim F_a \\   \bm h_{be} \sim F_b}} \left\{ R_a   + R_b  - R_e  \geq R_s \right\} \notag \\
 & \qquad \geq 1- \epsilon,   \label{eq:rrssrm-b}\\
 & \qquad {\rm Tr}(\bm Q_a) \leq P_a,   ~{\rm Tr}(\bm Q_b) \leq P_b, ~\bm Q_a \succeq \bm 0, ~ \bm Q_b\succeq \bm 0,    \label{eq:rrssrm-c}
\end{align}
\end{subequations}
where $R_a$, $R_b$ and $R_e$ are defined in \eqref{eq:def_Ra}-\eqref{eq:def_Re}; for notational convenience, we have dropped the arguments $(\bm Q_a, \bm Q_b)$ in $R_a$, $R_b$ and $R_e$;  $\epsilon \in (0,1)$ is the outage probability, specifying the chance of the sum secrecy rate falling below $R_s$ when $  \bm h_{ae}$ and $  \bm h_{be}$ vary randomly according to certain distributions $F_a$ and $F_b$, resp. Since the exact distributions of $  \bm h_{ae}$ and $ \bm h_{be}$ are not available, to be safe, the constraint \eqref{eq:rrssrm-b} requires that the secrecy  outage probability should be kept below $\epsilon$ for arbitrary distributions $F_a \in {\mathscr D}( {\bm \xi}_a,  {\bm \Omega}_a,  \tau_{a,1}, \tau_{a,2})$ and $F_b \in {\mathscr D}({\bm \xi}_b,  {\bm \Omega}_b,  \tau_{b,1}, \tau_{b,2})$.
The constraint~\eqref{eq:rrssrm-b} is meaningful in the sense of protecting transmission security and achieving maximal robustness against Eve's CSI uncertainty. However, from an optimization perspective, it  also poses a challenging task, since it involves infinite number of distributions. In the following, we  develop a tractable solution to problem~\eqref{eq:rrssrm} by leveraging the Lagrangian duality theory and the DC programming.

\subsection{A Tractable Safe Solution to the Robust SSRM Problem}

Since $ \bm h_{ae}$ and $  \bm h_{be}$ appear only in $R_e$, problem~\eqref{eq:rrssrm} can be recast as
 \begin{subequations} \label{eq:rssrm-eqv}
\begin{align}
& \max_{\bm Q_a, \bm Q_b, \nu_e}   ~ R_a + R_b - \log(1 + \nu_e) \label{eq:rssrm-eqv-a} \\
& {\rm s.t.}  ~\max_{\substack{  F_a \in \mathscr{D} (  {\bm \xi}_a,  {\bm \Omega}_a,  \tau_{a,1}, \tau_{a,2})\\   F_b \in \mathscr{D} (  {\bm \xi}_b,  {\bm \Omega}_b,  \tau_{b,1}, \tau_{b,2})}}  \mathbb{P}_{\substack{ \bm h_{ae} \sim F_a \\  \bm h_{be} \sim F_b}}  \left\{ \sum_{i\in \{a,b\}} \bm h_{ie}^H \bm Q_i \bm h_{ie}   \geq \sigma_e^2 \nu_e \right\} \notag \\
& \hspace{1cm} \leq \epsilon,   \label{eq:rssrm-eqv-b}\\
&   {\rm Tr}(\bm Q_a) \leq P_a,  ~{\rm Tr}(\bm Q_b) \leq P_b, ~ \bm Q_a \succeq \bm 0,~ \bm Q_b\succeq \bm 0,~ \nu_e \geq 0,  \label{eq:rssrm-eqv-c}
\end{align}
\end{subequations}
where we have  made a change of variable $\log(1 + \nu_e) = R_a + R_b - R_s$. Clearly, the main difficulty arises from \eqref{eq:rssrm-eqv-b}, which is a worst-case probabilistic constraint. In general, it is unlikely to express this worst-case probabilistic constraint into an explicit form. Herein, we focus on finding  a safe approximation to \eqref{eq:rssrm-eqv-b}, i.e., by replacing \eqref{eq:rssrm-eqv-b} with a relatively easy-to-handle constraint so that every  solution of the latter must fulfill~\eqref{eq:rssrm-eqv-b}. Following this idea, let us
summarize our main result on the safe approximation of \eqref{eq:rssrm-eqv} in the following theorem. The proof of the theorem is provided in  the next subsection.
\begin{Theorem} \label{theorm:1}
Consider the following problem:
  \begin{equation} \label{eq:rssr2-approx-eqv}
  \begin{aligned}
   &  \max    ~ R_a({\bm Q}_a, {\bm Q}_b) + R_b({\bm Q}_a,{\bm Q}_b) - \log(1 + \nu_e)  \\
&~~  {\rm s.t.}     ~   \sum_{i\in \{a,b\}} {\rm Tr} ( {\bm \Gamma}_i \bm \Psi_i +   {\bm \Phi}_i \bm \Xi_i) + {\alpha}_i \leq \epsilon  {\mu}, \\
&~~ ~ \begin{bmatrix}
   2{\bm B}_a &  &  {\bm \lambda}_a \\
   &  2{\bm B}_b   &    {\bm \lambda}_b \\
   {\bm \lambda}_a^H &  {\bm \lambda}_b^H & -( {\alpha}_a+ {\alpha}_b)
\end{bmatrix}
  \preceq \bm 0,  \\
    &~~~   \begin{bmatrix}
   2{\bm B}_a + \bm Q_a&  &  {\bm \lambda}_a \\
   &  2{\bm B}_b + \bm Q_b &    {\bm \lambda}_b \\
   {\bm \lambda}_a^H &  {\bm \lambda}_b^H & {\mu}- {\alpha}_a- {\alpha}_b - \sigma_e^2 \nu_e
\end{bmatrix}
 \preceq  \bm  0,  \\
& ~~ ~{\rm Tr}({\bm Q}_i) \leq P_i,  ~ {\bm Q}_i \succeq \bm 0,~  {\bm \Gamma}_i\succeq \bm 0, ~  {\bm \Phi}_i \succeq \bm 0, ~i\in \{a,b\}, \\
& ~~ ~ \bm \Gamma_i = \begin{bmatrix}
  \bm S_i & \bm \lambda_i \\
  \bm \lambda_i^H & \theta_i
\end{bmatrix},~ \bm \Phi_i = \begin{bmatrix}
  \bm A_i & \bm B_i \\
  \bm B_i & \bm C_i
\end{bmatrix},~i\in \{a,b\},  \\
& ~~ ~ {\mu} \geq 0 , ~~ \nu_e \geq 0, ~~\alpha_i\in \mathbb{R},~i\in \{a,b\},
  \end{aligned}
\end{equation}
where $(\{{\bm Q}_i, {\alpha}_i, { \bm \Gamma}_i,  {\bm \Phi}_i, \bm S_i, \bm \lambda_i, \theta_i, \bm A_i, \bm B_i, \bm C_i\}_{i\in \{a,b\}}, \nu_e,  {\mu})$ are optimization variables, and
\[ \bm \Psi_i \triangleq \begin{bmatrix}
 \tau_{i,1} \bm I &  -  {\bm \xi}_i \\
      -  {\bm \xi}_i^H & \tau_{i,1}
\end{bmatrix}, ~~ \bm \Xi_i \triangleq \begin{bmatrix}
    \tau_{i,2} \bm I & -   {  {\bm \Omega}}_i   \\
     -    {  {\bm \Omega}}_i    & \tau_{i,2}\bm I
  \end{bmatrix},~i\in\{a,b\} \] are constant matrices.
Then, problem~\eqref{eq:rssr2-approx-eqv} is a  safe approximation of the robust SSRM problem~\eqref{eq:rrssrm}; i.e., any  solution of problem~\eqref{eq:rssr2-approx-eqv} is a feasible solution of problem~\eqref{eq:rrssrm}.
\end{Theorem}

In~\eqref{eq:rssr2-approx-eqv}, the constraints are already convex and the objective can be written into a DC form:
\begin{equation*}
\begin{aligned}
  & R_a(\bm Q_a, \bm Q_b)+R_b(\bm Q_a, \bm Q_b)-\log(1+\nu_e)\\
   = & \varphi_1(\bm Q_a, \bm Q_b) - \varphi_2(\bm Q_a, \bm Q_b, \nu_e)
  \end{aligned}
\end{equation*}
where
\begin{align*}
\varphi_1(\bm Q_a, \bm Q_b)  \triangleq & \log( \sigma_a^2 + \zeta_a \bm h_{aa}^H \bm Q_a \bm h_{aa}+ \bm h_{ba}^H \bm Q_b \bm h_{ba}) \\
& + \log( \sigma_b^2 + \zeta_b \bm h_{bb}^H \bm Q_b \bm h_{bb}+ \bm h_{ab}^H \bm Q_a \bm h_{ab}) \\
\varphi_2(\bm Q_a, \bm Q_b, \nu_e) \triangleq & \log(1 + \nu_e) +  \log( \sigma_a^2 + \zeta_a \bm h_{aa}^H \bm Q_a \bm h_{aa}) \\
& + \log( \sigma_b^2 + \zeta_b \bm h_{bb}^H \bm Q_b \bm h_{bb}).
  \end{align*}
Therefore, again the DC programming can be employed to iteratively solve problem~\eqref{eq:rssr2-approx-eqv}. Moreover, by applying the classical DC convergence result~\cite[Theorem 1]{Luo}, we immediately conclude that every limit point generated by DC programm is a  stationary solution of problem~\eqref{eq:rssr2-approx-eqv}.

\subsection{Proof of Theorem~\ref{theorm:1}} \label{sec:rssrm_derive}

Our idea is to replace the    left-hand side (LHS) of \eqref{eq:rssrm-eqv-b} with its upper bound, so that after the replacement the resultant constraint is easier to handle than \eqref{eq:rssrm-eqv-b}. Since the LHS of \eqref{eq:rssrm-eqv-b} is the maximization of some probability measure, it can be upper bounded by its  dual optimal value by the Lagrangian duality theory. As such, the main endeavor in the sequel is  to   develop the dual of the maximization on the LHS of \eqref{eq:rssrm-eqv-b}.

To start, let ${\cal A} \triangleq \{( \bm h_{ae},  \bm h_{be})~ |~  \sum_{i \in \{a,b\}}  \bm h_{ie}^H \bm Q_i \bm h_{ie}   \geq \sigma_e^2 \nu_e \} $ and denote by $\mathbb{I}_{   {\cal A}}(  \bm h_{ae},   \bm h_{be})$ the indicator function over the set $\cal A$, i.e.,
\begin{equation}\label{eq:def_indicator}
 \mathbb{I}_{   {\cal A}}(  \bm h_{ae},   \bm h_{be}) = \left \{\begin{array}{l}
  1,~~{\rm if}~ (  \bm h_{ae},   \bm h_{be}) \in {\cal A}, \\
  0, ~~{\rm otherwise}.
\end{array} \right.
\end{equation}
Then, the outage probability in \eqref{eq:rssrm-eqv-b} can be  expressed as
\begin{equation}\label{eq:prob_cal}
\begin{aligned}
  & \mathbb{P}_{\substack{ \bm h_{ae} \sim F_a \\  \bm h_{be} \sim F_b}}  \left\{ \sum_{i\in \{a,b\}} \bm h_{ie}^H \bm Q_i \bm h_{ie}   \geq \sigma_e^2 \nu_e \right\}\\
  = & \mathbb{E} \left\{   \mathbb{I}_{   {\cal A}}(  \bm h_{ae},   \bm h_{be})  \right\} \\
   = & \int \int    \mathbb{I}_{   {\cal A}}(  \bm h_{ae},   \bm h_{be})  {\rm d}F_a(  \bm h_{ae}) {\rm d} F_b( \bm h_{be}).
\end{aligned}
\end{equation}
Recalling the definition  of $\mathscr{D}$ in \eqref{eq:def_D}, the maximization on the LHS of \eqref{eq:rssrm-eqv-b} can be written as the following constrained optimization problem w.r.t. the distributions $F_a$ and $F_b$:
\begin{subequations} \label{eq:rssrm-mu-prob}
  \begin{align}
    \max_{F_a, F_b}~ & \int \int    \mathbb{I}_{   {\cal A}}(  \bm h_{ae},   \bm h_{be})  {\rm d}F_a(  \bm h_{ae})  {\rm d} F_b(  \bm h_{be})   \label{eq:rssrm-mu-prob-a} \\
    {\rm s.t.}~ & \left \| \int   \bm h_{ie} {\rm d} F_i(  \bm h_{ie})   -   {\bm \xi}_{i} \right \|_2 \leq \tau_{i,1}, ~i\in \{a,b\} ,  \label{eq:rssrm-mu-prob-b} \\
    ~ &   \left \|  \int    \bm h_{ie}      \bm h_{ie}^H {\rm d} F_i(  \bm h_{ie})  - {  {\bm \Omega}}_i \right \|_2 \leq  \tau_{i,2},  ~i\in \{a,b\}, \label{eq:rssrm-mu-prob-c} \\
    ~ & \int {\rm d} F_i(  \bm h_{ie}) = 1, ~~i\in \{a,b\}.  \label{eq:rssrm-mu-prob-d}
  \end{align}
\end{subequations}

The constraint in~\eqref{eq:rssrm-mu-prob-b} is a second-order cone constraint, which can be reexpressed as the following linear matrix inequality (LMI) by using the Schur complement lemma~\cite[Lemma 4.2.1]{Bental}:
\begin{equation}\label{eq:rssrm2_mean_eqv}
\begin{aligned}
  & \int \begin{bmatrix}
    \tau_{i,1} \bm I &   \bm h_{ie} -  {\bm \xi}_i \\
    (  \bm h_{ie} -  {\bm \xi}_i)^H & \tau_{i,1}
  \end{bmatrix} {\rm d} F_i \succeq \bm 0\\
  \Longleftrightarrow & \int \begin{bmatrix}
     \bm 0&   \bm h_{ie}  \\
     \bm h_{ie}^H &  0
  \end{bmatrix} {\rm d} F_i(  \bm h_{ie}) \succeq   \begin{bmatrix}
    -\tau_{i,1} \bm I &     {\bm \xi}_i \\
      {\bm \xi}_i^H & - \tau_{i,1}
  \end{bmatrix}, ~~i\in \{a,b\}.
\end{aligned}
\end{equation}
Meanwhile, the spectral norm constraint in \eqref{eq:rssrm-mu-prob-c} can also be expressed as an LMI by noting the following equivalence~\cite{Bental}:
\[ \| \bm X \|_2 \leq \tau \Longleftrightarrow \begin{bmatrix}
  \tau \bm I_M & \bm X \\
  \bm X^H & \tau \bm I_N
\end{bmatrix} \succeq \bm 0 \]
for any matrix $\bm X\in \mathbb{C}^{M\times N}$ and $\tau \geq 0$.
Therefore, the constraint~\eqref{eq:rssrm-mu-prob-c} can be rewritten as
\begin{equation} \label{eq:rssrm2_cov_eqv}
\begin{aligned}
&  \int \begin{bmatrix}
    \tau_{i,2} \bm I &   \bm h_{ie}    \bm h_{ie} ^H -   {  {\bm \Omega}}_i   \\
     \bm h_{ie}    \bm h_{ie} ^H -    {  {\bm \Omega}}_i    & \tau_{i,2}\bm I
  \end{bmatrix} {\rm d}F_i(  \bm h_{ie}) \succeq \bm 0 \\
\Longleftrightarrow & \int \begin{bmatrix}
     \bm 0&   \bm h_{ie}    \bm h_{ie} ^H    \\
     \bm h_{ie}    \bm h_{ie} ^H     &  \bm 0
  \end{bmatrix}{\rm d}F_i(  \bm h_{ie}) \succeq
   \begin{bmatrix}
    -\tau_{i,2} \bm I &      {  {\bm \Omega}}_i   \\
        {  {\bm \Omega}}_i    &  - \tau_{i,2}\bm I
  \end{bmatrix}
  \end{aligned}
\end{equation}
for all $i\in \{a,b\}$. By substituting~\eqref{eq:rssrm2_mean_eqv} and \eqref{eq:rssrm2_cov_eqv} into \eqref{eq:rssrm-mu-prob}, problem~\eqref{eq:rssrm-mu-prob} can be equivalently written as
\begin{equation} \label{eq:rssr2-prob-eqv}
  \begin{aligned}
    \max_{F_a, F_b}~ ~& \int \int    \mathbb{I}_{   {\cal A}}(  \bm h_{ae}, \bm h_{be})  {\rm d}F_a(  \bm h_{ae})  {\rm d} F_b(  \bm h_{be})   \\
    {\rm s.t.}~ ~&  \eqref{eq:rssrm-mu-prob-d}, \eqref{eq:rssrm2_mean_eqv}, \eqref{eq:rssrm2_cov_eqv}~{\rm satisfied}.
  \end{aligned}
\end{equation}

Next, we derive the Lagrangian dual of problem~\eqref{eq:rssr2-prob-eqv}. Let
\[ \bar{\alpha}_i\in \mathbb{R}, ~~\bar{\bm \Gamma}_i = \begin{bmatrix}
  \bar{\bm S}_i & \bar{\bm \lambda}_i \\
  \bar{\bm \lambda}_i^H & \bar{\theta}_i
\end{bmatrix} \succeq \bm 0,~~{\rm  and }~~\bar{\bm \Phi}_i = \begin{bmatrix}
  \bar{\bm A}_i & \bar{\bm B}_i \\
  \bar{\bm B}_i & \bar{\bm C}_i
\end{bmatrix} \succeq \bm 0
 \] for $i\in \{a,b\}$ be the Lagrangian multipliers associated with \eqref{eq:rssrm-mu-prob-d}, \eqref{eq:rssrm2_mean_eqv} and \eqref{eq:rssrm2_cov_eqv}, respectively. Then, the Lagrangian of problem~\eqref{eq:rssr2-prob-eqv} can be expressed as
\begin{equation}
\begin{aligned}
  {\cal L} = & \int \int    \mathbb{I}_{   {\cal A}}(  \bm h_{ae}, \bm h_{be})  {\rm d}F_a(  \bm h_{ae})  {\rm d} F_b(  \bm h_{be}) \\
  & + \sum_{i\in \{a,b\}} \bar{\alpha}_i( 1 - \int {\rm d} F_i(  \bm h_{ie}) )  \\
    & +  \sum_{i \in \{a, b\}} \int \bar{\bm \Gamma}_i \bullet \begin{bmatrix}
    \bm 0  &   \bm h_{ie}   \\
    \bm h_{ie}^H & 0
  \end{bmatrix}   {\rm d} F_i(  \bm h_{ie})  + \sum_{i \in \{a, b\} } \bar{\bm \Gamma}_i \bullet \bm \Psi_i  \\
   & +   \sum_{i \in \{a, b\}} \int \bar{\bm \Phi}_i \bullet \begin{bmatrix}
      \bm 0 &   \bm h_{ie}    \bm h_{ie}^H  \\
     \bm h_{ie}    \bm h_{ie}^H     &  \bm 0
  \end{bmatrix}  {\rm d}F_i(  \bm h_{ie}) \\
  & + \sum_{i\in \{a,b\}} \bar{\bm \Phi}_i \bullet \bm \Xi_i  \\
  = & \int \int    \Big\{ \mathbb{I}_{   {\cal A}}(  \bm h_{ae}, \bm h_{be})  +  \sum_{i\in \{a,b\}} \left( - \bar{\alpha}_i+  2{\rm Re}\{ \bar{\bm \lambda}_i^H    \bm h_{ie} \}  \right.    \\
  & \left.  + 2 \bm h_{ie}^H \bar{\bm B}_i   \bm h_{ie}   \right ) \Big\}    {\rm d}F_a(  \bm h_{ae})  {\rm d} F_b(  \bm h_{be}) \\
  & +   \sum_{i \in \{a,b\}} (  \bar{\bm \Gamma}_i \bullet \bm \Psi_i + \bar{\bm \Phi}_i \bullet \bm \Xi_i + \bar{\alpha}_i)
  \end{aligned}
\end{equation}
where for notational simplicity, we have denoted $\bm A \bullet \bm B \triangleq  {\rm Tr}(\bm A^H \bm B)$ and
\[ \bm \Psi_i \triangleq \begin{bmatrix}
 \tau_{i,1} \bm I &  -  {\bm \xi}_i \\
      -  {\bm \xi}_i^H & \tau_{i,1}
\end{bmatrix}, ~~ \bm \Xi_i \triangleq \begin{bmatrix}
    \tau_{i,2} \bm I & -   {  {\bm \Omega}}_i   \\
     -    {  {\bm \Omega}}_i    & \tau_{i,2}\bm I
  \end{bmatrix},~i\in\{a,b\}. \]
Therefore, the dual of problem~\eqref{eq:rssr2-prob-eqv} is given by
\begin{subequations} \label{eq:rssr2-prob-dual}
  \begin{align}
    & \min_{\{\bar{\alpha}_i, \bar{\bm \Gamma}_i, \bar{\bm \Phi}_i\}_{i\in \{a,b\}}}~~   \sum_{i \in \{a,b\}}   \bm \bar{\bm \Gamma}_i \bullet \bm \Psi_i + \bar{\bm \Phi}_i \bullet \bm \Xi_i + \bar{\alpha}_i \label{eq:rssr2-prob-dual-a}  \\
    &~ {\rm s.t.}~~    \mathbb{I}_{\cal A}(  \bm h_{ae},   \bm h_{be}) + \sum_{i\in\{a,b\}} 2 {\rm Re}\{ \bar{\bm \lambda}_i^H    \bm h_{ie} \}  \notag\\
    & \quad \qquad + 2    \bm h_{ie}^H \bar{\bm B}_i   \bm h_{ie} - \bar{\alpha}_i  \leq 0, ~~\forall~   \bm h_{ae},  \bm h_{be} \in \mathbb{C}^N, \label{eq:rssr2-prob-dual-b} \\
    &  \quad \qquad  \bar{\bm \Gamma}_i \succeq \bm 0, \quad \bar{\bm \Phi}_i \succeq \bm 0, ~~i\in \{a,b\}.
  \end{align}
\end{subequations}

To express the constraint~\eqref{eq:rssr2-prob-dual-b} into a more tractable form, let us define
\[\bar{\bm B} = 2{\rm Diag}(\bar{\bm B}_a,~ \bar{\bm B}_b),~ ~\bar{\bm \lambda} = [\bar{\bm \lambda}_a^T, ~\bar{\bm \lambda}_b^T]^T,~~  \bm h_e =  [ \bm h_{ae}^T,  ~ \bm h_{be}^T ]^T.\]
Recalling the definition of $\mathbb{I}_{\cal A}(  \bm h_{ae},   \bm h_{be})$ in~\eqref{eq:def_indicator},  it is clear that \eqref{eq:rssr2-prob-dual-b} holds if and only if
\hspace{-2cm} \begin{subnumcases}{\label{eq:rssr2-dual-const-eqv}}
\hspace{-5pt} \begin{bmatrix}
    \bm h_e \\
  1
\end{bmatrix}^H\hspace{-5pt}
 \begin{bmatrix}
  \bar{\bm B} & \bar{\bm \lambda} \\
  \bar{\bm \lambda}^H & -(\bar{\alpha}_a+ \bar{\alpha}_b)
\end{bmatrix}\hspace{-5pt}
\begin{bmatrix}
    \bm h_e \\
  1
\end{bmatrix}\hspace{-2pt} \leq 0,  \forall   \bm h_e \in \mathbb{C}^N \label{eq:rssr2-dual-const-eqva}\\
 \hspace{-5pt} \begin{bmatrix}
    \bm h_e \\
  1
\end{bmatrix}^H \hspace{-5pt}\begin{bmatrix}
  \bar{\bm B} & \bar{\bm \lambda} \\
  \bar{\bm \lambda}^H & -(\bar{\alpha}_a+ \bar{\alpha}_b-1)
\end{bmatrix}
\hspace{-5pt}
\begin{bmatrix}
    \bm h_e \\
  1
\end{bmatrix} \hspace{-4pt} \leq 0 , \forall   \bm h_e \in {\cal A}  \label{eq:rssr2-dual-const-eqvb}
\end{subnumcases}
For~\eqref{eq:rssr2-dual-const-eqva}, the quadratic inequality holds for any vector $\bm h_e\in \mathbb{C}^N$ if and only if
\begin{equation} \label{eq:dual-LMI1}
   \begin{bmatrix}
  \bar{\bm B} & \bar{\bm \lambda} \\
  \bar{\bm \lambda}^H & -(\bar{\alpha}_a+\bar{\alpha}_b)
\end{bmatrix}
  \preceq \bm 0.
\end{equation}
For~\eqref{eq:rssr2-dual-const-eqvb}, it amounts to the following implication:
\begin{equation} \label{eq:rssr2-dual-const-eqvb-LMI}
\begin{aligned}
&    \begin{bmatrix}
    \bm h_e \\
  1
\end{bmatrix}^H   \begin{bmatrix}
    \bm Q & \bm 0 \\
    \bm 0^H & -\sigma_e^2 \nu_e
\end{bmatrix}
  \begin{bmatrix}
    \bm h_e \\
  1
\end{bmatrix} \geq 0  \Longrightarrow     \\
&   \begin{bmatrix}
    \bm h_e \\
  1
\end{bmatrix}^H \begin{bmatrix}
  \bar{\bm B} & \bar{\bm \lambda} \\
  \bar{\bm \lambda}^H & -(\bar{\alpha}_a+\bar{\alpha}_b-1)
\end{bmatrix}
\begin{bmatrix}
    \bm h_e \\
  1
\end{bmatrix} \leq 0
\end{aligned}
\end{equation}
for any $\bm h_e$, where $\bm Q \triangleq {\rm Diag}(\bm Q_a, ~\bm Q_b)$. Since both sides of the above implication are quadratic  w.r.t. $\bm h_e$, it follows from the $\cal S$-procedure~\cite{Bental} that the implication \eqref{eq:rssr2-dual-const-eqvb-LMI} holds if and only if the following   matrix inequality is true:
\begin{equation} \label{eq:dual-LMI2}
\begin{bmatrix}
  \bar{\bm B} & \bar{\bm \lambda} \\
  \bar{\bm \lambda}^H & 1-\bar{\alpha}_a-\bar{\alpha}_b
\end{bmatrix}
+ \bar{\mu} \begin{bmatrix}
  \bm Q & \bm 0 \\
  \bm 0^H  &   - \sigma_e^2 \nu_e
\end{bmatrix}
\preceq  \bm  0
\end{equation}
for some slack variable $\bar{\mu} \geq 0$.

Now, by substituting \eqref{eq:dual-LMI1} and \eqref{eq:dual-LMI2} into~\eqref{eq:rssr2-prob-dual}, the dual of problem~\eqref{eq:rssr2-prob-eqv} can be written into a more compact form:
\begin{subequations} \label{eq:rssr2-dual}
  \begin{align}
     \min_{\bar{\mu}, \{\bar{\alpha}_i, \bar{\bm \Gamma}_i, \bar{\bm \Phi}_i\} } & ~    \sum_{i \in \{a,b\}}   \bar{\bm \Gamma}_i \bullet \bm \Psi_i + \bar{\bm \Phi}_i \bullet \bm \Xi_i + \bar{\alpha}_i \label{eq:rssr2-dual-a}  \\
     {\rm s.t.} & ~       \begin{bmatrix}
  \bar{\bm B} & \bar{\bm \lambda} \\
  \bar{\bm \lambda}^H & -(\bar{\alpha}_a+ \bar{\alpha}_b)
\end{bmatrix}
  \preceq \bm 0,  \label{eq:rssr2-dual-b} \\
    & ~  \begin{bmatrix}
  \bar{\bm B} & \bar{\bm \lambda} \\
  \bar{\bm \lambda}^H & 1-\bar{\alpha}_a-\bar{\alpha}_b
\end{bmatrix}
+ \bar{\mu} \begin{bmatrix}
  \bm Q & \bm 0 \\
  \bm 0^H  &   - \sigma_e^2 \nu_e
\end{bmatrix}
\preceq  \bm  0, \label{eq:rssr2-dual-c} \\
& ~ \bar{\bm \Gamma}_i\succeq \bm 0, ~\bar{\bm \Phi}_i \succeq \bm 0, i\in\{a,b\}, ~\bar{\mu} \geq 0. \label{eq:rssr2-dual-d}
  \end{align}
\end{subequations}
From the duality theory, we know that the optimal value of problem~\eqref{eq:rssr2-prob-eqv} or equally the LHS of \eqref{eq:rssrm-eqv-b} is upper bounded by the optimal value of problem~\eqref{eq:rssr2-dual}. Hence, we arrive at the following key result:
\begin{Claim}\label{prop:rssrm2}
  The outage probability constraint~\eqref{eq:rssrm-eqv-b} holds if the optimal value of problem~\eqref{eq:rssr2-dual} is no greater than $\epsilon$.
\end{Claim}
We should mention that Claim~\ref{prop:rssrm2} states only a sufficient, but not necessary condition for fulfilling the constraint~\eqref{eq:rssrm-eqv-b}. This is because there may exist duality gap between problem~\eqref{eq:rssr2-prob-eqv} and its dual \eqref{eq:rssr2-dual}. Also notice that the optimal value of \eqref{eq:rssr2-dual} is no greater than $\epsilon$ if and only if there exists some feasible point of \eqref{eq:rssr2-dual} such that the corresponding objective value in \eqref{eq:rssr2-dual-a}  is no greater than $\epsilon$. Consequently, a safe approximation of problem~\eqref{eq:rrssrm} is readily obtained by replacing the LHS of \eqref{eq:rssrm-eqv-b} with  \eqref{eq:rssr2-dual-a} and adding \eqref{eq:rssr2-dual-b}-\eqref{eq:rssr2-dual-d} into the constraints of \eqref{eq:rssrm-eqv}, viz.,
\begin{subequations} \label{eq:rssr2-approx}
  \begin{align}
   &  \max_{ \substack{\{\bm Q_i,\bar{\alpha}_i, \bar{\bm \Gamma}_i, \bar{\bm \Phi}_i\}_{i} \\ \nu_e, \bar{\mu}  }}    ~ R_a(\bm Q_a, \bm Q_b) + R_b(\bm Q_a, \bm Q_b) - \log(1 + \nu_e) \label{eq:rssr2-approx-a} \\
&~~  {\rm s.t.}     ~   \sum_{i\in \{a,b\}} {\rm Tr} (\bar{\bm \Gamma}_i \bm \Psi_i +  \bar{\bm \Phi}_i \bm \Xi_i) + \bar{\alpha}_i \leq \epsilon,  \label{eq:rssr2-approx-b}\\
&~~ ~ \begin{bmatrix}
  \bar{\bm B} & \bar{\bm \lambda} \\
 \bar{ \bm \lambda}^H & -(\bar{\alpha}_a+ \bar{\alpha}_b)
\end{bmatrix}
  \preceq \bm 0,  \label{eq:rssr2-approx-c} \\
    &~~~   \begin{bmatrix}
  \bar{\bm B} & \bar{\bm \lambda} \\
  \bar{\bm \lambda}^H & 1-\bar{\alpha}_a-\bar{\alpha}_b
\end{bmatrix}
+ \bar{\mu} \begin{bmatrix}
  \bm Q & \bm 0 \\
  \bm 0^H  &   - \sigma_e^2 \nu_e
\end{bmatrix}
\preceq  \bm  0, \label{eq:rssr2-approx-d} \\
& ~~ ~{\rm Tr}(\bm Q_i) \leq P_i,   \bm Q_i \succeq \bm 0, \bar{\bm \Gamma}_i\succeq \bm 0,  \bar{\bm \Phi}_i \succeq \bm 0, ~i\in \{a,b\}, \label{eq:rssr2-approx-e}\\
& ~~~ \bar{\mu} \geq 0 , ~~ \nu_e \geq 0. \label{eq:rssr2-approx-f}
  \end{align}
\end{subequations}
Notice that all the constraints in \eqref{eq:rssr2-approx} are  convex w.r.t. the optimization variables except for~\eqref{eq:rssr2-approx-d}, where $\bar{\mu}$ is coupled with $\bm Q$ and $\nu_e$. To turn~\eqref{eq:rssr2-approx-d} into a convex constraint, we make use of the  following observation:
\begin{Observation}\label{claim:positve_mu}
Any feasible $\bar{\mu}$ of problem~\eqref{eq:rssr2-approx} must be strictly positive.
\end{Observation}
The proof of Observation~\ref{claim:positve_mu} can be found in Appendix~\ref{appendix:claim_strict_positive}.

Since $\bar{\mu}>0$, we  make the following change of variables:
\[  {\bm \Phi}_i = \bar{\bm \Phi}_i/ \bar{\mu}, ~~  {\bm \Gamma}_i = \bar{\bm \Gamma}_i/ \bar{\mu},~ ~ {\alpha}_i = \bar{\alpha}_i/\bar{\mu},~~ {\mu} =1/\bar{\mu},\]
and reexpress problem~\eqref{eq:rssr2-approx}  as problem~\eqref{eq:rssr2-approx-eqv}. This completes the proof of Theorem~\ref{theorm:1}.

\section{Numerical Results} \label{sec:sim}
In this section, we use Monte Carlo simulations to evaluate the performances of the proposed designs under both perfect and imperfect CSI cases. We will first consider the perfect
CSI case in the first subsection, and then the imperfect CSI case
in the second subsection.

\subsection{The Perfect CSI Case} \label{sec:sim_perfect_CSI}

The results to be presented in this subsection are based on
the following simulation settings, unless otherwise specified: The number of transmit antennas at Alice and Bob are $N=4$;
all the receive noises have  zero mean and unit variance. For simplicity, we assume that Alice and Bob have the same SI residual factor $\zeta_a=\zeta_b=\zeta = 0.01$, and the same transmit power $P_a=P_b=P=5$~dB. All the channels were randomly generated following i.i.d. complex Gaussian distribution with zero mean and unit variance.

\begin{figure}[!htp]
  \centering
  \includegraphics[width=2.8in]{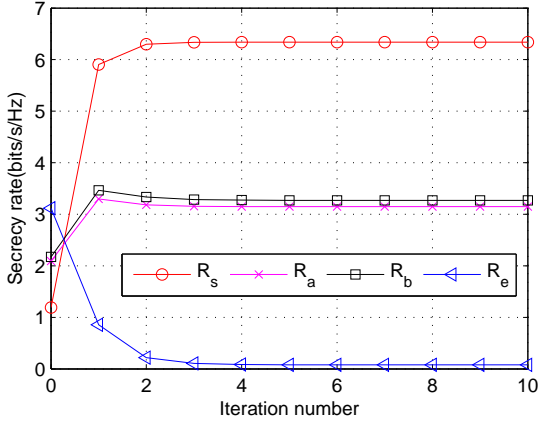}
  \caption{secrecy rate vs. iteration number}\label{fig:rate_iter}
  \vspace{-5pt}
\end{figure}

Fig.~\ref{fig:rate_iter} shows convergence behaviors of the ADC algorithm for one randomly generated problem instance. From the figure we see that the ADC algorithm converges quickly  within 3 iterations. After the first two iterations, Eve's rate can be largely suppressed  from the initial three bits/s/Hz to nearly zero, and meanwhile, Alice and Bob's rates are improved.

\begin{figure}[!htp]
  \centering
  \includegraphics[width=2.8in]{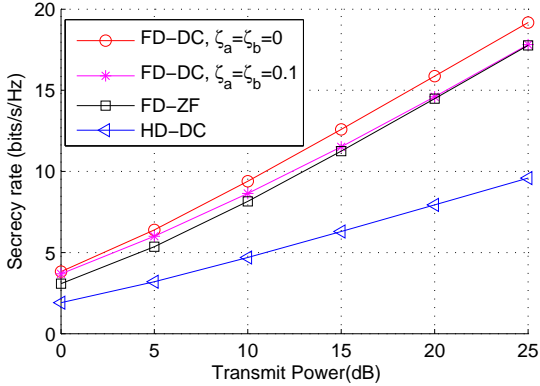}
  \caption{Secrecy rate vs. transmit power at Alice and Bob}\label{fig:rate_pow}
  \vspace{-5pt}
\end{figure}

Fig.~\ref{fig:rate_pow} plots the secrecy rate against the transmit power $P$ at Alice and Bob under various designs. In the legend, ``FD-DC'' represents the proposed ADC approach, cf.~Algorithm~\ref{algorithm:1}; ``HD-DC'' represents the conventional half-duplex (HD) bidirectional transmissions, which suffers from a rate reduction by half.  ``FD-ZF'' is similar to ``FD-DC'', except that the transmit covariance matrices $\bm Q_a$ and $\bm Q_b$ are enforced in the null space of their respective SI channels, so that the residual SI can be completely eliminated at the cost of one spatial degree of freedom (d.o.f.) loss per transmitter. Owing to the zero-forcing (ZF) constraints, the subproblems in~\eqref{eq:AO_a} and \eqref{eq:AO_b}
are degenerated into the secrecy design problem for the MISOME wiretap channel in~\cite{Khisti}, and the optimal solution can be obtained through generalized eigenvalue decomposition; readers are referred to~\cite{Khisti} for details. From the figure we see that FD-DC attains the best rate performance among the three designs, and that FD transmission is in general better than HD. Moreover, with the increase of the SI residual factor $\zeta$, FD-DC gets closer to FD-ZF, especially for large transmit power. This is because large $\zeta$ and $P$ may incur significant SI, which would offset the d.o.f. gain of FD-DC.

\begin{figure}[!htp]
  \centering
  \includegraphics[width=2.8in]{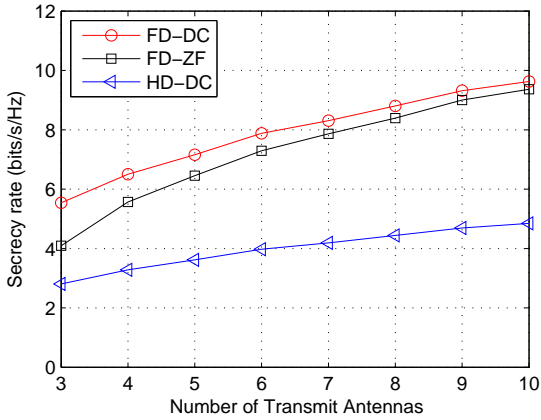}
  \caption{Secrecy rate vs. number of transmit antennas at Alice and Bob}\label{fig:rate_ant}
  \vspace{-5pt}
\end{figure}

Fig.~\ref{fig:rate_ant} shows the relationship between the sum secrecy rates and the number of transmit antennas  for different designs.
From Fig.~\ref{fig:rate_ant}, we see that with the increase of the transmit  antennas, the secrecy rates of all methods increase. Specifically, for small  number of antennas, there are notable rate gaps between FD-DC and the other two designs. For large number of antennas, FD-ZF is able to approach FD-DC, because of the negligible spatial d.o.f. loss by ZF. In addition, HD-DC is far inferior to the other two FD-based designs, owing to the half reduction of the rate induced by the HD transmissions.

\begin{figure}[!htp]
\begin{center}
\subfigure[][]{\resizebox{.25\textwidth}{!}{\includegraphics{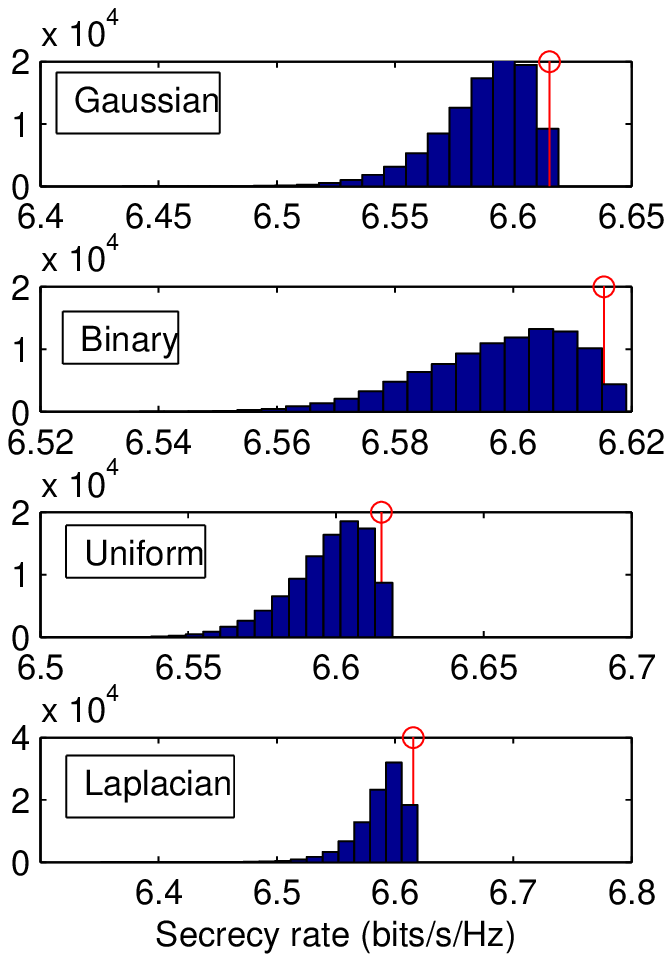}}}
\subfigure[][]{\resizebox{.228\textwidth}{!}{\includegraphics{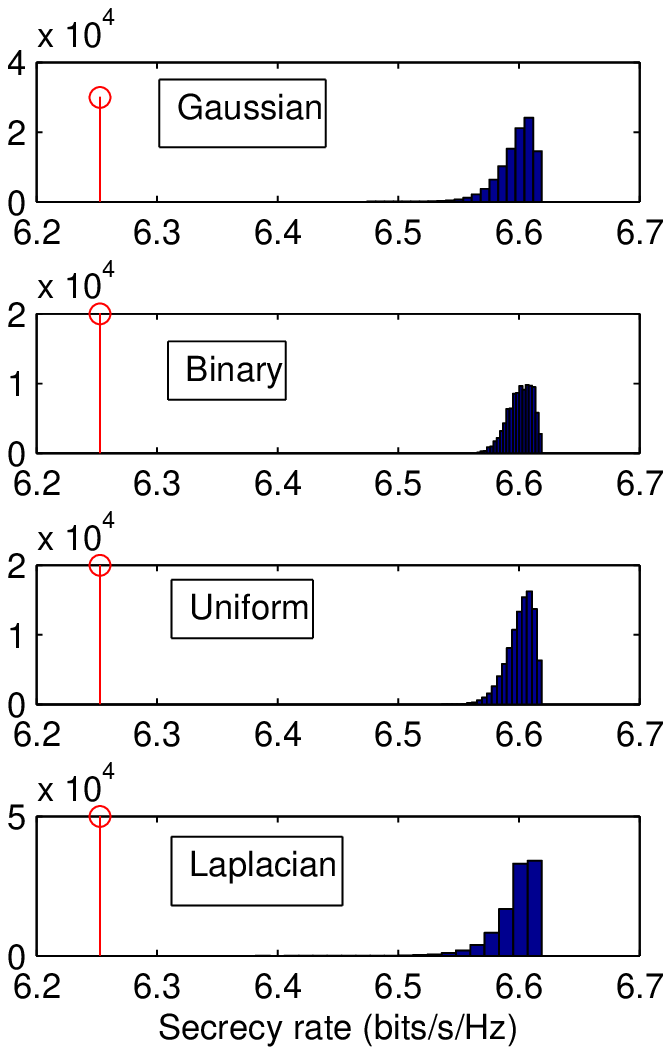}}}
\end{center}
\caption{Secrecy rate histograms  under different distributions ($\tau_1=\tau_2 = 0, {\rho}=0.002$) (a) Nonrobust ADC design with estimated mean, and (b) Robust DC design with exact moment.}
\label{fig:rob1_hist}
\end{figure}

\begin{figure}[!htp]
  \centering
  \includegraphics[width=2.8in]{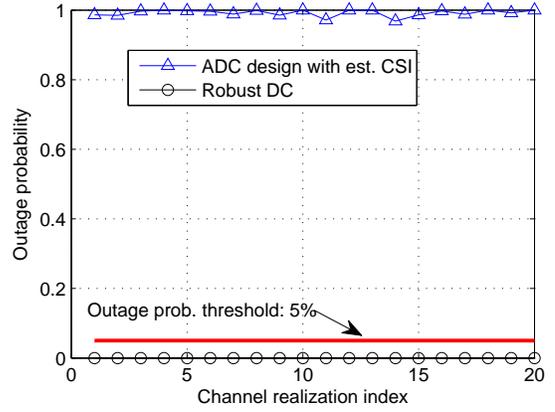}
  \caption{Empirical outage probability of the nonrobust ADC and the robust DC for the first 20 channel realizations}\label{fig:rob1_prob_ch}
  \vspace{-5pt}
\end{figure}

\begin{figure}[!htp]
\begin{center}
\subfigure[][]{\resizebox{.24\textwidth}{!}{\includegraphics{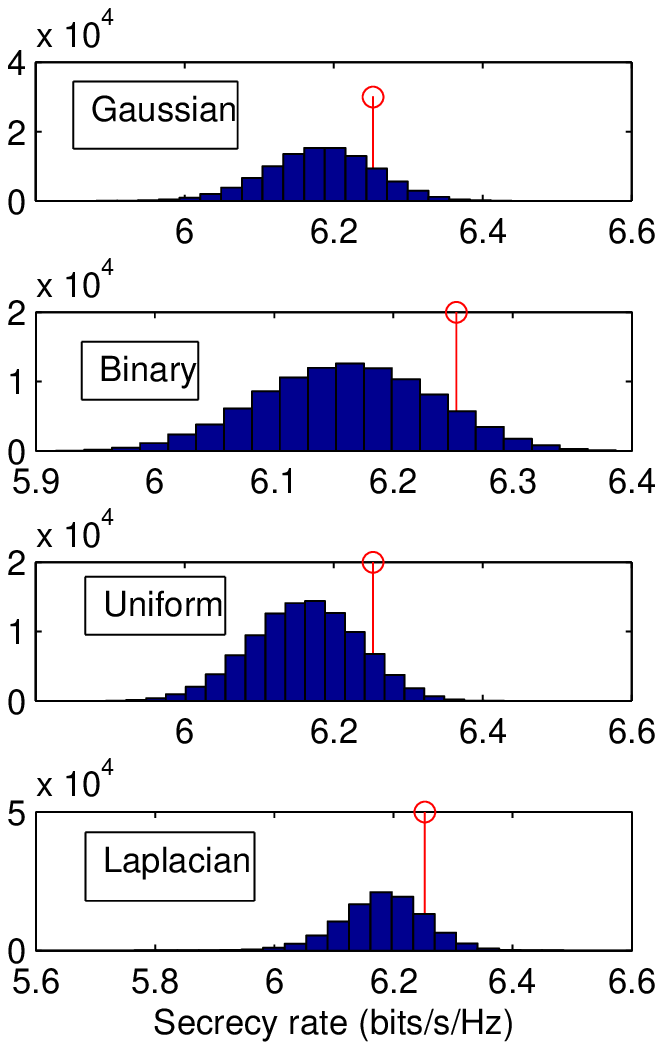}}}
\subfigure[][]{\resizebox{.235\textwidth}{!}{\includegraphics{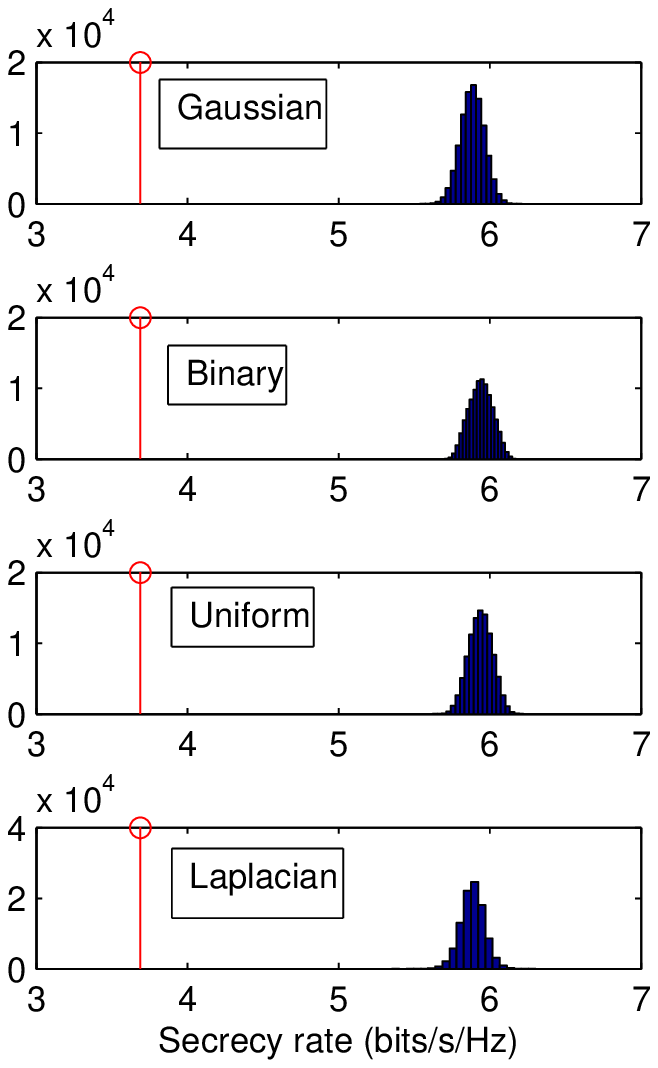}}}
\end{center}
\caption{Secrecy rate histograms  under different distributions ($\tau_1=\tau_2 = 0.05$, $ {\rho} =0.002$) (a) Robust DC based on estimated moments, and (b) Robust DC with moment uncertainty.}
\label{fig:rob2_hist}
\end{figure}

\begin{figure}[!htp]
  \centering
  \includegraphics[width=2.8in]{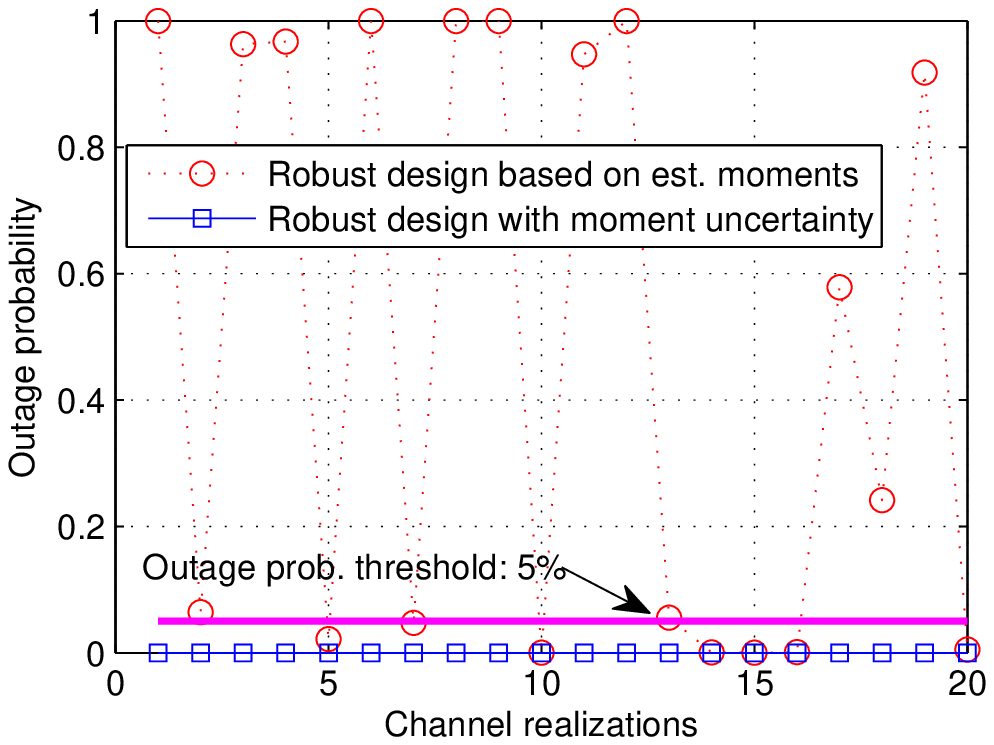}
  \caption{Empirical outage probability of the robust DCs with and without considering moment uncertainty for the first 20 channel realizations}\label{fig:rob2_prob_ch}
  \vspace{-5pt}
\end{figure}

\subsection{The Imperfect CSI Case} \label{sec:sim_imperfect_CSI}
In the next two examples, we consider the imperfect CSI case. The simulation settings are  basically identical to those of the perfect CSI case, i.e., $N=4$, $\sigma_a^2=\sigma_b^2=\sigma_e^2=1$, $\zeta_a=\zeta_b=\zeta=0.01$ and $P_a=P_b=P=5$~dB. The secrecy outage probability threshold is set to $\epsilon = 5\%$. As for the CSI uncertainty model, we set the estimated mean $\bm \xi_a= \bm \xi_b=\bm \xi =  0.01\times(\bm 1_N+\sqrt{-1} \bm 1_N)$, the estimated second-order moment $\bm \Omega_a= \bm \Omega_b = \bm \Omega = \bm \xi \bm \xi^H + \rho \bm I_{N}$, and $\tau_{i,1}=\tau_{1}$, $\tau_{i,2}=\tau_{2}, i\in \{a,b\}$ for simplicity.

{\it Example 1) Exact moment case:} We first test the robustness of the proposed robust DC design under the exact moment case via setting $\tau_{1}=\tau_2 = 0$. For comparison, we also run the ADC Algorithm~\ref{algorithm:1} with the estimated mean; i.e., by  replacing $\bm h_{ie}, i\in \{a,b\}$ in problems~\eqref{eq:AO_a} and \eqref{eq:AO_b} with the estimated mean $\bm \xi_i,  i\in \{a,b\}$. After convergence, we obtain some secrecy rate value $R_s^{ADC}$ and the transmit solution $(\bm Q_{a}^{ADC}, \bm Q_b^{ADC})$. Then, we use the solution $(\bm Q_{a}^{ADC}, \bm Q_b^{ADC})$ to perform the subsequent tests.
Since it is unlikely to generate all possible distributions in $\mathcal{D}(\bm \xi, \bm \Omega, 0, 0)$,  we consider four representative distributions, namely,  Gaussian distribution, binary distribution, uniform distribution and Laplace distribution. Our test is conducted as follows: Suppose that we have obtained a transmit solution $(\bm Q_a, \bm Q_b)$ by either ADC or robust DC. Then, we randomly generate $10^5$ Eve's channel realizations  for each of the four distributions, and calculate the secrecy rate $R_{s}^{j,t}(\bm Q_a, \bm Q_b)$ for the $t$th channel realization of the type  $j=1,\ldots, 4$ distribution. After obtaining all $R_{s}^{j,t}(\bm Q_a, \bm Q_b)$ for each $j$, we plot the histograms of  $R_{s}^{j,t}(\bm Q_a, \bm Q_b)$ for $j=1,\ldots,4$.
Fig.~\ref{fig:rob1_hist}(a) shows the histograms of $R_{s}^{j,t}(\bm Q_a^{ADC}, \bm Q_b^{ADC})$, where the red circled line corresponds to the secrecy rate value $R_s^{ADC}$ obtained from Algorithm~\ref{algorithm:1}. From the figure, we see that for most channel realizations, the empirical secrecy rate is below $R_s^{ADC}$; that is, the ADC design suffers from severe secrecy outage, and  cannot provide satisfactory transmission security because of ignoring CSI imperfection.
Fig.~\ref{fig:rob1_hist}(b) shows  the result of the robust DC. Clearly,  all the empirical secrecy rates of the robust DC are above the secrecy rate threshold obtained from~\eqref{eq:rssr2-approx-eqv}, and thus no secrecy outage happens under the four tested distributions.  Fig.~\ref{fig:rob1_hist}(a) and (b) are obtained from one randomly generated problem instance. To further evaluate the robustness, we repeated the above test for 20 randomly generated problem instances, and plotted the worst outage probability (among the four distributions) of the two designs in Fig.~\ref{fig:rob1_prob_ch}. As seen, the outage probability of the ADC algorithm  is far higher than the desired threshold $5\%$, whereas the robust DC is always below the threshold.

{\it Example 2) Uncertain moment case:} In this example, we test the performance of the robust DC under moment uncertainty via setting $\tau_1=\tau_2=0.05$. Similar to Example 1, we compare two robust DC designs: One is based on the estimated moment information without considering the moment  uncertainty, i.e., by presuming $\tau_1=\tau_2 = 0$; the other takes into account the moment uncertainty. Fig.~\ref{fig:rob2_hist} shows the histograms of the two robust designs under different distributions. As seen, by slightly  perturbing  the first and the second-order moments, the performance of the robust DC design without considering moment uncertainty degrades significantly --- for most channel realizations, secrecy outage happens. This is in sharp contrast to the previous example in Fig.~\ref{fig:rob1_hist}(b), where no secrecy outage happens when the moment information is accurately estimated. For the robust DC  with considering moment uncertainty in Fig.~\ref{fig:rob2_hist}(b),  its empirical  secrecy  rates are all above the rate threshold, thereby no secrecy outage occurring. Similar to Fig.~\ref{fig:rob1_prob_ch}, we also tested the outage probability  of the two robust DCs for 20 randomly generated problem instances and the result is shown in Fig.~\ref{fig:rob2_prob_ch}. We see that the robust DC without considering moment uncertainty has only  seven problem instances with secrecy outage probability $\leq 5\%$, whereas the design with considering moment uncertainty keeps the secrecy outage probability below $5\%$.

\section{Conclusion}\label{sec:conclusion}
This paper has considered  transmit covariances optimization for maximizing the sum secrecy rate (SSR) of a full-duplex bidirectional communication system. Both perfect and imperfect CSI of the eavesdropper are considered. For the former, we developed an alternating DC (ADC)   programming approach  to maximizing the sum secrecy rate, and established its  stationary solution convergence. For the latter, we developed a tractable safe approximation to the robust SSR maximization by leveraging the Lagrangian duality theory and the DC programming. Simulation results demonstrated that our proposed designs can achieve a better secrecy rate performance and  provide desired robustness against CSI imperfection.



\appendix

\subsection{Proof of Proposition~\ref{lemma:closed_form_sol}}\label{appendix:closed_form_sol}
The inner maximization in \eqref{eq:dual} amounts to
\begin{equation*}
  \max_{\bm W_a\succeq \bm 0} ~ \log(1 + \hat{\bm h}_{ab}^H \bm W_a \hat{\bm h}_{ab} ) - {\rm Tr}((\bm M_a+ \lambda_a \bm I) \bm W_a).
\end{equation*}
By making a change of variable
\begin{equation}\label{eq:var_change}
  \breve{\bm W}_a = (\bm M_a+ \lambda_a \bm I)^{1/2} \bm W_a (\bm M_a+ \lambda_a \bm I)^{1/2},
\end{equation}
the above maximization can be further rewritten as
\begin{equation} \label{eq:adc_sol}
  \max_{\breve{\bm W}_a\succeq \bm 0} ~ \log(1 + \breve{\bm h}_{ab}^H \breve{\bm W}_a \breve{\bm h}_{ab} ) - {\rm Tr}(\breve{\bm W}_a),
\end{equation}
where $\breve{\bm h}_{ab} \triangleq (\bm M_a+ \lambda_a \bm I)^{-1/2} \hat{\bm h}_{ab}$. Note that
\[ \breve{\bm h}_{ab}^H \breve{\bm W}_a \breve{\bm h}_{ab}\leq {\rm Tr}(\breve{\bm W}_a) \| \breve{\bm h}_{ab} \|^2. \]
Therefore, 
\begin{align*}
  & \log(1 + \breve{\bm h}_{ab}^H \breve{\bm W}_a \breve{\bm h}_{ab} ) - {\rm Tr}(\breve{\bm W}_a) \\
   \leq & \log(1 + {\rm Tr}(\breve{\bm W}_a) \| \breve{\bm h}_{ab} \|^2 ) - {\rm Tr}(\breve{\bm W}_a) \\
   \leq & \max_{{\rm Tr}(\breve{\bm W}_a ) \geq 0}~ \log(1 + {\rm Tr}(\breve{\bm W}_a) \| \breve{\bm h}_{ab} \|^2 ) - {\rm Tr}(\breve{\bm W}_a)\\
   = & \log\big(1 + [\|  \breve{\bm h}_{ab} \|^2-1]^+\big) - \big[1- 1/\|  \breve{\bm h}_{ab} \|^2 \big]^+,
\end{align*}
where  the last equality holds when ${\rm Tr}(\breve{\bm W}_a ) =  \big[1- 1/\|  \breve{\bm h}_{ab} \|^2 \big]^+$. Clearly, the above upper bound is achieved if and only if
\begin{equation} \label{eq:opt_sol_inner}
   \breve{\bm W}_a =  \frac{\big[1- 1/\|  \breve{\bm h}_{ab} \|^2 \big]^+}{ \|\breve{\bm h}_{ab}\|^2}  \breve{\bm h}_{ab} \breve{\bm h}_{ab}^H
 \end{equation}
in \eqref{eq:adc_sol}. That is, \eqref{eq:opt_sol_inner} gives the optimal solution for problem~\eqref{eq:adc_sol}, or equivalently optimal $\bm W_a$ for the inner maximization in~\eqref{eq:dual} via the relation \eqref{eq:var_change}.

Next, we derive the upper bound on the dual optimal $\lambda_a^\star$. Since $\lambda_a^\star >0$ implies ${\rm Tr}(\bm W_a^\star) = P_a$, we must have $\kappa_a>0$ and thus
$\| (\bm M_a + \lambda_a^\star \bm I)^{-1/2} \hat{\bm h}_{ab}\|^2>1$. Notice that $\| (\bm M_a + \lambda_a^\star \bm I)^{-1/2} \hat{\bm h}_{ab}\|^2 = \hat{\bm h}_{ab}^H (\bm M_a + \lambda_a^\star \bm I)^{-1}  \hat{\bm h}_{ab} \leq \hat{\bm h}_{ab}^H   (\lambda_a^\star \bm I)^{-1}  \hat{\bm h}_{ab} = (\lambda_a^\star)^{-1} \| \hat{\bm h}_{ab}\|^2$. Therefore, we arrive at $(\lambda_a^\star)^{-1} \| \hat{\bm h}_{ab}\|^2>1$, i.e., $\lambda_a^\star< \| \hat{\bm h}_{ab}\|^2$. Moreover, it is known that ${\rm Tr}\left(\bm W_a^\star \right)$ is a nonincreasing function w.r.t. $\lambda_a^\star$ \cite[Sec. 13.1, Lemma 1]{DG2008}, i.e., the larger $\lambda_a^\star$ leads to the smaller ${\rm Tr}\left(\bm W_a^\star \right)$, and vice versa. Hence, the optimal $\lambda_a^\star$ can be efficiently computed via bisection over the interval $(0,~ \| \hat{\bm h}_{ab}\|^2)$ such that  the complementarity condition is satisfied.

\subsection{Proof of Proposition~\ref{prop:adc_convergence}} \label{appendix:adc_convergence}
Since problems~\eqref{eq:dc_a} and \eqref{eq:dc_b} both have compact feasible sets, the iterates $\{ (\bm W_a^k, \bm W_b^k)\}_k$ generated by Algorithm~\ref{algorithm:1} must be bounded. By the Weierstrass theorem  $\{ (\bm W_a^k, \bm W_b^k)\}_k$ contains at least one limit point.  To establish the second part of the proposition, we show that   Algorithm~\ref{algorithm:1} is a special case of the BSUM algorithm in~\cite{Luo}, and  the convergence result of the latter applies.
For self-containedness, let us first briefly review the BSUM algorithm. Consider the following maximization problem\footnote{The original BSUM  considers a minimization problem; herein we changed  the minimization as maximization  to fit into our  context.}:
\begin{equation}\label{eq:bsum_main}
\begin{aligned}
  \max_{\bm x \in \mathbb{R}^m}& ~ g(\bm x)  \\
  {\rm s.t.}& ~ \bm x \in  \mathcal{X} \triangleq  \mathcal{X}_1 \times \ldots \times  \mathcal{X}_n,
\end{aligned}
\end{equation}
where $\bm x =  (\bm x_1,\ldots, \bm x_n)$, $\sum_{i=1}^n m_i = m$ and $  \mathcal{X}_i \subseteq \mathbb{R}^{m_i}$ is the feasible set for the $i$th block variable. The BSUM algorithm for~\eqref{eq:bsum_main} is summarized in the following procedure.
\begin{algorithm}
\begin{algorithmic}[1]
  \State Find a feasible point $\bm x^0\in \mathcal{X}$ and set $k=0$
   \Repeat
   \State $k=k+1$, $i=(r ~{\rm mod} ~n)+1$
\State Let \begin{equation} \label{eq:bsum_subprob}
  \mathcal{X}^k  = \arg\max_{\bm x_i \in \mathcal{X}_i} u_i(\bm x_i, \bm x^{k-1})
\end{equation}
\State Set $\bm x_i^k$ to be an arbitrary element in $\mathcal{X}^k$
\State Set $\bm x_\ell^k = \bm x_\ell^{k-1},~\forall~\ell \neq i$
\Until{some stopping criterion is met}
\end{algorithmic}
\end{algorithm}

In line 4, the function   $u_i: \mathbb{R}^{m_i} \mapsto \mathbb{R}$ is a surrogate function used for updating the $i$th block variable. In particular, $u_i(\cdot,\cdot)$ satisfies the following key properties:
\begin{enumerate}
  \item $u_i(\bm y_i; \bm y ) = g(\bm y), ~~\forall~\bm y\in \mathcal{X}, ~\forall~i$,
  \item $u_i(\bm x_i; \bm y ) \leq g(\bm y_1,\ldots, \bm y_{i-1}, \bm x_i, \bm y_{i+1},\ldots, \bm y_{n}),~\forall~\bm x_i \in {\cal X}_i, ~\forall~\bm y\in \mathcal{X}, ~\forall~i$,
      \item $u_i^\prime(\bm x_i, \bm y; \bm d_i) = g^\prime(\bm y;\bm d), ~\forall~\bm d= (\bm 0, \ldots, \bm d_i, \ldots, \bm 0)$ s.t. $\bm y_i + \bm d_i \in \mathcal{X}_i,~\forall~i$, where $u_i^\prime$ and $g^\prime$ denote the directional derivatives of $u_i$ and $g$ along the directions $\bm d_i$ and $\bm d$, resp.,
          \item  $u_i(\bm x_i, \bm y)$ is continuous in $(\bm x_i, \bm y), ~\forall ~i$.
\end{enumerate}
The following convergence result was established in \cite[Theorem 2]{Luo}.
\begin{Theorem}[\hspace{-0.1pt}\cite{Luo}] \label{theorem:bsum_conv}
Suppose that the function $u_i(\bm x_i, \bm y)$ is quasi-concave in $\bm x_i$ for
$i = 1, \ldots,  n$. Furthermore, assume that the subproblem~\eqref{eq:bsum_subprob}
has a unique solution for any point $\bm x^{k-1} \in \mathcal{X}$. Then, every limit point $\bm z$ of the iterates
generated by the BSUM algorithm is a coordinatewise minimum of \eqref{eq:bsum_main}. In addition,
if $g(\cdot)$ is regular\footnote{If $g$ is differentiable, then it is automatically regular.} at $\bm z$, then $\bm z$ is a stationary point of~\eqref{eq:bsum_main}.
\end{Theorem}

Now, we establish the second part of the proposition by verifying that the ADC algorithm fulfills the conditions required by Theorem~\ref{theorem:bsum_conv}.
First of all, since  $f_a(\bm W_a; \bm W_a^k, \bm W_b^k)$ is obtained by keeping the concave part $\tilde{R}_a(\bm W_a, \bm W_b^k)$ and linearizing the convex part $\tilde{R}_b(\bm W_a, \bm W_b^k)- \tilde{R}_e(\bm W_a, \bm W_b^k)$. It can be easily verified that $f_a(\bm W_a; \bm W_a^k, \bm W_b^k)$ is concave and satisfies the following properties:
\begin{enumerate}
  \item $f_a(\bm W_a; \bm W_a^k, \bm W_b^k) \leq \tilde{R}_a(\bm W_a, \bm W_b^k)+\tilde{R}_b(\bm W_a, \bm W_b^k)- \tilde{R}_e(\bm W_a, \bm W_b^k) $ for all feasible $\bm W_a$,
  \item  $f_a(\bm W_a^k; \bm W_a^k, \bm W_b^k) = \tilde{R}_a(\bm W_a^k, \bm W_b^k)+\tilde{R}_b(\bm W_a^k, \bm W_b^k)- \tilde{R}_e(\bm W_a^k, \bm W_b^k) $,
  \item $\nabla_{\bm W_a} f_a(\bm W_a^k; \bm W_a^k, \bm W_b^k) = \nabla_{\bm W_a}[\tilde{R}_a(\bm W_a^k, \bm W_b^k)+\tilde{R}_b(\bm W_a^k, \bm W_b^k)- \tilde{R}_e(\bm W_a^k, \bm W_b^k)] $,
      \item $f_a(\bm W_a; \bm W_a^k, \bm W_b^k)$ is continuous in $(\bm W_a; \bm W_a^k, \bm W_b^k)$.
\end{enumerate}
That is, $f_a(\bm W_a; \bm W_a^k, \bm W_b^k)$ is a tight lower bound on $\tilde{R}_a(\bm W_a, \bm W_b^k)+\tilde{R}_b(\bm W_a, \bm W_b^k)- \tilde{R}_e(\bm W_a, \bm W_b^k)$.

Next, we show that the subproblem~\eqref{eq:dc_a} has a unique optimal solution.  In Proposition~\ref{lemma:closed_form_sol} and \ref{lemma:closed_form_sol_C2}, we have derived the optimal solution $\bm W_a^\star$, which must be of rank one. Suppose on the contrary that there are two distinct rank-one optimal solutions,  denoted  by $\bm W_{a,1}^\star$ and $\bm W_{a,2}^\star$ for problem~\eqref{eq:dc_a}. Then, their convex combination $\tilde{\bm W}_a = \rho \bm W_{a,1}^\star + (1-\rho) \bm W_{a,2}^\star$ for any $\rho\in (0,~1)$ must be optimal for problem~\eqref{eq:dc_a}. Let us consider the following two cases:
\begin{enumerate}
  \item $\bm W_{a,1}^\star \neq c \bm W_{a,2}^\star$ for any $c \geq 0$, i.e., $\bm W_{a,1}^\star$ and $\bm W_{a,2}^\star$ are linearly independent. Then, their convex combination  $\tilde{\bm W}_a$  must be of rank two, which contradicts with the rank-one solution structure revealed in Proposition~\ref{lemma:closed_form_sol} and \ref{lemma:closed_form_sol_C2}.

  \item $\bm W_{a,1}^\star = c \bm W_{a,2}^\star$ for some $c \geq 0$ and $c\neq 1$, i.e., $\bm W_{a,1}^\star$ and $\bm W_{a,2}^\star$ are linearly dependent with different scaling. Then, $\tilde{\bm W}_a = ((c-1)\rho + 1) \bm W_{a,2}^\star$. Substituting $\tilde{\bm W}_a$ into the objective of  \eqref{eq:dc_a_eqv} yields
       \begin{align*}
         g(\rho) \triangleq & \log \left( 1 + ((c-1)\rho + 1)\hat{\bm h}_{ab}^H  \bm W_{a,2}^\star  \hat{\bm h}_{ab} \right) \\
         & - \big((c-1)\rho + 1\big){\rm Tr}(\bm M_a  \bm W_{a,2}^\star ),
       \end{align*}
      which is strictly concave w.r.t. $\rho$ (for fixed $\bm W_{a,2}^\star$). Notice that both $g(0)$ and $g(1)$ correspond to the optimal value of \eqref{eq:dc_a_eqv}. Hence,
      \begin{equation} \label{eq:rank-1-obj}
        g(\rho) \leq g(0) = g(1),~~ \forall~\rho\in (0,~1).
      \end{equation}
      On the other hand, it follows from the strict concavity of $g(\rho)$ that
      \[g(\rho) > (1-\rho)g(0) + \rho g(1) = g(0), ~~\forall~\rho \in(0,~1),\]
      which contradicts with~\eqref{eq:rank-1-obj}.
\end{enumerate}
Therefore, we conclude that the optimal solution for problem~\eqref{eq:dc_a} must be unique. Similarly, one can  verify that the  tight lower bound property and the solution uniqueness also  hold for   $f_b(\bm W_b; \bm W_a^{k+1},\bm W_b^k)$. Therefore, the ADC Algorithm~\ref{algorithm:1} fulfills all the conditions of  Theorem~\ref{theorem:bsum_conv}, and consequently  Proposition~\ref{prop:adc_convergence} holds.

\subsection{Proof of Proposition~\ref{prop:lemma_multieve_eqv}}\label{appendix:lemma_multieve_eqv}
For ease of exposition, let us define
\[ \psi(\bm W_a, \bm \gamma) \triangleq     \log(1 + \hat{\bm h}_{ab}^H \bm W_a \hat{\bm h}_{ab} ) - \sum_{i=1}^I \gamma_i {\rm Tr}(\bm M_{i} \bm W_a).\]
From the definition of $g(\bm \gamma)$, it holds that
\[ g(\bm \gamma) = \psi(\bm W_a^\star(\bm \gamma), \bm \gamma),\]
where $\bm W_a^\star(\bm \gamma)$ is an optimal solution of problem~\eqref{eq:multieve_eqv_b}. Therefore, for any feasible $\gamma$ and $\bar{\bm \gamma}$ we have
 \begin{equation*}
  \begin{aligned}
    g(\bar{\bm \gamma}) & = \psi(\bm W_a^\star(\bar{\bm \gamma}), \bar{\bm \gamma}) \\
    & \geq  \psi(\bm W_a^\star( {\bm \gamma}), \bar{\bm \gamma}) \\
    & =  \psi(\bm W_a^\star( {\bm \gamma}),  {\bm \gamma}) - \textstyle \sum_{i=1}^I {\rm Tr}(\bm M_i \bm W_a^\star( {\bm \gamma}))^T( \bar{\gamma}_i - \gamma_i )\\
    & = g( {\bm \gamma})  - \textstyle \sum_{i=1}^I {\rm Tr}(\bm M_i \bm W_a^\star( {\bm \gamma})) ( \bar{\gamma}_i - \gamma_i ).
  \end{aligned}
\end{equation*}
That is, $-[{\rm Tr}(\bm M_1 \bm W_a^\star( {\bm \gamma})), \ldots, {\rm Tr}(\bm M_I \bm W_a^\star( {\bm \gamma}))]$ is a subgradient of $g(\bm \gamma)$ at the point $\bm \gamma$. Moreover, given $\bm \gamma$, problem~\eqref{eq:multieve_eqv_b} takes a similar form as \eqref{eq:dc_a_eqv} by treating $\sum_{i=1}^I \gamma_i \bm M_i$ as $\bm M_a$ in \eqref{eq:dc_a_eqv}. Hence, the optimal $\bm W_a^\star(\bm \gamma)$ of problem~\eqref{eq:multieve_eqv_b} must be of rank one and unique according to the uniqueness proof in Appendix~\ref{appendix:adc_convergence}. Now, it follows from Danskin's theorem~\cite{Bertsekas} that  $g(\bm \gamma)$  is differentiable, and the subgradient $-[{\rm Tr}(\bm M_1 \bm W_a^\star( {\bm \gamma})), \ldots, {\rm Tr}(\bm M_I \bm W_a^\star( {\bm \gamma}))]$ is automatically the gradient of $g(\bm \gamma)$ at $\bm \gamma$.

To prove the second part, notice that
\begin{align*}
  &\min_{i=1,\ldots, I} \log(1 + \hat{\bm h}_{ab}^H \bm W_a \hat{\bm h}_{ab} ) - {\rm Tr}(\bm M_{i} \bm W_a)\\
 = & \min_{\gamma_i\geq 0, \forall i,~ \bm \gamma^T \bm 1 = 1} \log(1 + \hat{\bm h}_{ab}^H \bm W_a \hat{\bm h}_{ab} ) - \sum_{i=1}^I \gamma_i {\rm Tr}(\bm M_{i} \bm W_a), \\
 =  & \min_{\gamma_i\geq 0, \forall i,~ \bm \gamma^T \bm 1 = 1}  \psi(\bm W_a, \bm \gamma).
\end{align*}
Problem~\eqref{eq:dc_sub_multi_Eve} can be rewritten as
\begin{equation}\label{eq:max_min_saddle}
\max_{ {\bm W_a \succeq \bm 0,  {\rm Tr}(\bm W_a) \leq P_a}}   ~  \min_{ {\gamma_i\geq 0, \forall i,  \bm \gamma^T \bm 1 = 1}} \psi(\bm W_a, \bm \gamma).
\end{equation}
Since $\bm W_a$ and $\bm \gamma$ lie in the compact sets, and $\psi(\bm W_a, \bm \gamma)$ is convex in $\bm \gamma$ and concave in $\bm W_a$, it follows from Sion's max-min theorem~\cite{Sion58} that  the maximization and the minimization in~\eqref{eq:max_min_saddle} can be exchanged without sacrificing optimality. Thus, problem~\eqref{eq:max_min_saddle} is equivalent to
  \begin{equation} \label{eq:dc_multi_Eve_eqv}
   \min_{ {\gamma_i\geq 0, \forall i, \bm \gamma^T \bm 1 = 1}} ~\left\{ \max_{ {\bm W_a \succeq \bm 0, {\rm Tr}(\bm W_a) \leq P_a}}   \psi(\bm W_a, \bm \gamma) \right\}.
\end{equation}
Using the definition of $g(\bm \gamma)$ yields the desired result in~\eqref{eq:multieve_eqv_a}.

%
%
%
%

\subsection{Proof of Observation~\ref{claim:positve_mu}}\label{appendix:claim_strict_positive}
Notice that $\bm \Psi_i$ and $\bm \Xi_i$ are both positive semidefinite; hence, $\sum_{i\in \{a,b\}}  {\rm Tr} ( \bar{\bm \Gamma}_i \bm \Psi_i +  \bar{\bm \Phi}_i \bm \Xi_i) \geq 0$, which together with~\eqref{eq:rssr2-approx-b} implies that \begin{equation}\label{eq:appendix_claim}
  \bar{\alpha}_a+ \bar{\alpha}_b \leq \epsilon <1.
\end{equation}
Let us suppose on the contrary $\bar{\mu} =0$. It follows from~\eqref{eq:rssr2-approx-d} that $1-\bar{\alpha}_a- \bar{\alpha}_b \leq 0$, i.e.,
\[
  \bar{\alpha}_a+\bar{\alpha}_b \geq 1,
\]
which contradicts with~\eqref{eq:appendix_claim}.

\end{document}